\documentclass[apj,numberedappendix]{emulateapj}

\def\gtorder{\mathrel{\raise.3ex\hbox{$>$}\mkern-14mu
             \lower0.6ex\hbox{$\sim$}}}
\def\ltorder{\mathrel{\raise.3ex\hbox{$<$}\mkern-14mu
             \lower0.6ex\hbox{$\sim$}}}

\slugcomment{Submitted to ApJ}

\shorttitle{Radio transients}

\shortauthors{Ofek et al.}

\begin{document}

\title{A VLA search for 5\,GHz radio transients and variables at low Galactic latitudes}
\author{E.~O.~Ofek\altaffilmark{1}$^{,}$\altaffilmark{2},
D.~A.~Frail\altaffilmark{3},
B.~Breslauer\altaffilmark{3}$^{,}$\altaffilmark{4},
S.~R.~Kulkarni\altaffilmark{1},
P.~Chandra\altaffilmark{5},
A.~Gal-Yam\altaffilmark{6},
M.~M.~Kasliwal\altaffilmark{1},
and N.~Gehrels\altaffilmark{7}}

\altaffiltext{1}{Division of Physics, Mathematics and Astronomy, California Institute of Technology, Pasadena, CA 91125, USA}
\altaffiltext{2}{Einstein fellow}
\altaffiltext{3}{National Radio Astronomy Observatory, P.O. Box O, Socorro, NM 87801}
\altaffiltext{4}{Department of Physics and Astronomy, Oberlin College, Oberlin, Ohio 44074-1088}
\altaffiltext{5}{Department of Physics, Royal Military College of Canada, Kingston, ON, Canada}
\altaffiltext{6}{Benoziyo Center for Astrophysics, Weizmann Institute of Science, 76100 Rehovot, Israel}
\altaffiltext{7}{NASA-Goddard Space Flight Center, Greenbelt, Maryland 20771}

\begin{abstract}


We present the results of a 5\,GHz survey with the Very Large Array (VLA) and
the expanded VLA, designed to search for short-lived ($\ltorder1$\,day)
transients and to characterize the variability of radio sources at milli-Jansky
levels. A total sky area of 2.66\,deg$^{2}$, spread over 141 fields at low
Galactic latitudes ($b\cong6$--$8$\,deg) was observed 16 times with a cadence
that was chosen to sample timescales of days, months and years. Most of the
data were reduced, analyzed and searched for transients in near real time.
Interesting candidates were followed up using visible light telescopes
(typical delays of 1--2\,hr) and the X-Ray Telescope on board the {\it Swift}
satellite. The final processing of the data revealed a single possible
transient with a flux density of $f_{\nu}\cong2.4$\,mJy. This implies a
transients sky surface density of
$\kappa(f_{\nu}>1.8\,{\rm mJy})=0.039_{-0.032,-0.038}^{+0.13,+0.18}\,{\rm deg}^{-2}$
(1, 2-$\sigma$ confidence errors). This areal density is consistent with the
sky surface density of transients from the Bower et al. survey extrapolated
to 1.8\,mJy. Our observed transient areal density is consistent with a
Neutron Stars (NSs) origin for these events. Furthermore, we use the data to
measure the sources variability on days to years time scales, and we present
the variability structure function of 5\,GHz sources. The mean
structure function
shows a fast increase on $\approx1$\,day time scale, followed by a slower
increase on time scales of up to 10\,days. On time scales between
10--60\,days the structure function is roughly constant. We find that
$\gtorder 30$\% of the unresolved sources brighter than 1.8\,mJy are variable
at the $>4$-$\sigma$ confidence level, presumably due mainly to refractive
scintillation.

\end{abstract}


\section{Introduction}
\label{sec:Introduction}

Radio surveys of the sky in the time domain have often been used to
identify new astrophysical phenomena. Highly variable radio sources
can serve as signposts to compact, high energy objects which are
accompanied by high magnetic fields and/or relativistic particle
acceleration. Radio variability from quasars and $\gamma$-ray bursts
(Dent 1965; Frail et al. 1997) was used to infer bulk relativistic
motions in these objects (Rees 1967; Goodman et al. 1987). Notable new
phenomena identified from radio time-domain surveys include the
discovery of the first pulsars (Hewish et al. 1968), the Galactic
high-energy binary LSI+61$^\circ$303 (Gregory and Taylor 1978), the
anomalous variability of 4C\,21.53 that lead to the discovery of
millisecond pulsars (Backer et al. 1982),
and the still-mysterious
extreme scattering events (Fiedler et al. 1987).

More recent surveys have found several new types of radio transients
whose identity has remained unknown or not well understood (e.g., Hyman et al. 2005;
McLaughlin et al. 2006; Bower et al.  2007; Lorimer et al.  2007;
Niinuma et al.  2007; Kida et al. 2008; Matsumura et al. 2009).

Specifically, Bower et al.
(2007) re-analyzed 944 epochs of Very Large Array\footnote{The Very
Large Array is operated by the National Radio Astronomy Observatory (NRAO),
a facility of the National Science Foundation operated under
cooperative agreement by Associated Universities, Inc.} (VLA)
observations, taken about once per week for twenty two years, of a
single calibration field.  These authors discovered a total of ten
transients, eight in the 5-GHz band and two in the 8-GHz band. Eight
of these transients were detected in a single epoch.
Therefore, their duration is shorter than the
time between successive epochs (one week) and longer than the exposure
time (20\,min).
Moreover, the
majority of these sources do not have any optical counterpart coinciding
with their position.
The lack of optical counterparts down to limiting
magnitudes of 27.6 in $g$-band and 26.5 in $R$-band is especially puzzling
and significantly limits the classes of objects that can be associated
with these events (Ofek et al. 2010).

In a possibly related work, Kuniyoshi et al. (2006), Niinuma et al.
(2007), and Kida et al. (2008) reported a search for radio
transients using an East-West interferometer of the Nasu Pulsar
Observatory (located in Tochigi Prefecture, Japan) of Waseda
University.  To date, this program reported 11 bright radio transients
with flux densities above 1\,Jy in the 1.4-GHz band.

Recently, in Ofek et al. (2010) we suggested that the properties of
the single epoch ``Bower et al. transients'' and the Nasu transients
are consistent with emerging from a single
class of objects, namely isolated old Neutron Stars (NS).
Specifically, the NS hypothesis is consistent with the rate,
energetics, sky surface density, source number count function and the
lack of optical counterparts.

In this paper we present a new VLA survey for radio transients and
variables at low Galactic latitudes.  Our main motivation for this
survey was to detect more examples of this new class of short-lived
radio transients, with the goal of identifying them in real-time in
order to find their counterparts at other wavelengths for further
study. A second, and equally important motivation for this survey was
to characterize the transient and variable
radio sky with a sensitivity and cadence
which had not been carried out previously.

The organization of this paper is as follows.
In \S\ref{Prev} we provide a summary of previous radio transient and
variability surveys.
In \S\ref{SurveyObs} we present the observations,
while the data reduction is outlined in \S\ref{Red}.
The results from our real time transients search 
are provided in \S\ref{RealTimeTran}.
\S\ref{PostSurveyCat} present the source catalogs generated in the post
survey phase.
The final post survey transient search is described
in \S\ref{PostSurveyTran} while the sources variability study is
presented in \S\ref{Variability}.
The implications of this study are discussed in \S\ref{Disc}
and we summarize in \S\ref{Sum}.
In addition three appendices discussing:
flux calibration;
the statistics of max/min of a time series;
and transient areal density calculation in the case of
a beam with non-uniform sensitivity
are provided.

\section{Previous GHz Surveys for Transients and Variables}
\label{Prev}

Existing 0.8-8\,GHz surveys have already explored,
to some extent, the dynamic radio
sky with a wide range of sensitivities, angular resolution and
cadences. 
However, compared with synoptic surveys at higher frequencies 
(infra-red to $\gamma$-rays)
the radio sky remains poorly explored.
In Table~\ref{Tab:PreviousSurveys} we summarize
past synoptic radio surveys.
For each survey we list also the number of transients, as well as variables
which vary by more than 50\%.
We note however, that comparison of these numbers is complicated
due to several factors.
A radio image
may be accomplished either through a single pointing,
or adding several scans taken at different times.
If the time span, $\delta{t}$, containing
all the observation composing a single ``epoch''
is larger than the transient duration
(or variability time scale)
then the survey sensitivity to transients (variables) is degraded.
Additionally, the probability of 
detecting significant variability
depends on $\delta{t}$ and the typical time scale
between epochs ($\Delta{t}$), through the variability structure function.
Depending on the statistical method used to define
the variability amplitude, it may also affected  by the
number of epochs ($N_{{\rm ep}}$) in the survey.
\begin{deluxetable*}{llllllllllll}
\tablecolumns{12}
\tabletypesize{\scriptsize}
\tablecaption{Previous GHz Transient and variability Surveys}
\tablehead{
\colhead{$\nu$}            &
\colhead{Area}             &
\colhead{Direction}        &
\colhead{$\Delta{\theta}$} & 
\colhead{N$_{{\rm ep}}$}    & 
\colhead{$\delta{t}$}      & 
\colhead{$\Delta{t}$}      & 
\colhead{rms}              & 
\colhead{Sources}          & 
\colhead{Tran.}            &
\colhead{Var.}             &
\colhead{Ref.}             \\
\colhead{GHz}      &
\colhead{deg$^{2}$} &
\colhead{deg}      &
\colhead{$''$}     & 
\colhead{}         & 
\colhead{}         & 
\colhead{}         & 
\colhead{mJy}      & 
\colhead{}         & 
\colhead{}         &
\colhead{}         &
\colhead{}
}
\startdata
0.84 &2776   & $\delta<-30$    & $\sim45$   &2\tablenotemark{a} & 12\,hr     & 1\,day--20\,yr    & 2.8   & 29730   & 15        & $\sim10$  & [14]       \\ 
1.4  & 0.22  & $l=150$, $b=+53$& 4.5                 & 3        & 6\,hrs     & 19\,d, 17\,m      & 0.015 &\nodata  & 0         & $2\%$     & [1]        \\ 
1.4  & 2.6   & $l=151$, $b=+24$& 60                  & 16       & 12\,hrs    & 1-12\,d, 1-3\,m   & 0.7   & 245     & 0         & $\sim1\%$ & [2]        \\ 
1.4  & 120   & S. Galactic Cap & 5                   & 2        & days       & 7\,yr             & 0.15  & 9086    & 0         &    1.4\%  & [3]        \\ 
1.4  & 2500  & $b\gtorder30$   & 45                  & 2        & days       & $\sim$years       & 0.45  & 7181    & 1         &\nodata    & [5,6,7]    \\ 
1.4  & 2870\tablenotemark{b}& $+32>\delta>+42$&$24'\times2.4'$&$\sim1000$&4\,min  &1\,d          & 300   &\nodata  &$11$       &\nodata    & [8,9,10,11]\\ 
1.4  & 0.2   & $l=57$, $b=+81$ & 20                  & 1852     & minutes    & 1\,day--23\,yr    & 2     & 10      & 0         &\nodata    & [19]       \\ 
1.4  & 690   & $l=70$, $b=+64$ & 150                 & 2        & months     & 15\,yr            & 3.94  & 4408    & 0         &$\sim0.1\%$& [4]        \\ 
1.4  & 690   & $l=70$, $b=+64$ & 150                 & 12       & $>1$\,day  & days--months      & 38    & 4408    & 0         &$\ltorder0.5\%$& [20]   \\ 
1.4  & 0.2   & phase calib.    & \nodata             & 151      & $5$\,min   & days-years        & $\sim1$&\nodata & 0         & \nodata   & [21]       \\ 
3.1  & 10    & $l=57$, $b=+67$ & 100                 & 2        &months      &15\,yr             & 0.25  & 425     & 1\tablenotemark{c}&\nodata& [12]   \\ 
4.9  & 0.1   & phase calib.    & \nodata&$\sim390$\tablenotemark{d}&$5$\,min & days-years        & $\sim1$&\nodata & 0         & \nodata   & [21]       \\ 
4.9  & 0.69  & Extragalactic   & 0.5-15              & 2        &60\,min     & 1-100d            & 0.05  & \nodata & 0         & \nodata   & [15]       \\ 
4.9  & 23.2  &$\vert b\vert<0.4$& 5                  & 3        &90\,s       & 2\,m--15\,yr      & 0.2   &2700     & 0         & 15        & [16]       \\ 
4.9  & 500   & $\vert b\vert<2$& 180                 & 16       &2\,min      & 1\,day--5\,yr     & 4.6   & 1274    & 1         & $\sim0.5$ & [18]       \\ 
4.9  & 19924 & $75>\delta>0$   & 210                 & 2        & $\sim$week & 1\,yr             & 5     & 75162   & 0         & $>40$     & [17]       \\ 
4.9  & 0.07  & $l=115$, $b=+36$& 5                   & 626      &20\,mim     & 1\,week--22\,yr   & 0.05  & 8       & 7\tablenotemark{e}& 0 & [13]       \\ 
8.5  & 0.02  & $l=115$, $b=+36$& 3                   & 599      &20\,min     & 1\,week--22y      & 0.05  & 4       & 1\tablenotemark{f}& 0 & [13]       \\ 
8.5  & 0.04  & phase calib.    & \nodata&$\sim308$\tablenotemark{g}&$5$\,min & days-years        & $\sim1$&\nodata & 0         & \nodata   & [21]       \\ 
\hline
4.9  & 2.6   & $\vert b\vert\approx7$ & 15           & 16       &50\,s       & 1\,d--2\,yr       & 0.15  &$\sim200$& 1         &$0.3-30\%$& This paper    
\enddata
\tablecomments{{\it Columns description:} $\Delta{\theta}$ is the beam full width at half power; $N_{{\rm ep}}$ is the number of epochs;
$\delta{t}$ is the time span over which each epoch was obtained (see text);
$\Delta{t}$ is the range of time separations between epochs;
Sources is the number of persistent sources detected;
Tran. is the number of transients found by the survey;
Var. is the number or percentage of variables showing variability larger than 50\%.
We note that strong variables are defined differently by each survey.
Therefore, these numbers provide only a qualitative comparison
between the surveys.
References: [1] Carilli et al.~(2003), [2] Frail et al. (1994), [3] de Vries et al. (2004), [4] Croft et al. (2010), [5] Levinson et al. (2002), [6] Gal-Yam et al. (2006), [7] Ofek et al. (2010), [8] Matsumura et al. (2009), [9] Kida et al. (2008), [10] Kuniyoshi et al. (2007), [11] Matsumura et al. (2007), [12] Bower et al. (2010), [13] Bower et al. (2007), [14] Bannister et al. (2010), [15] Frail et al. (2003), [16] Becker et al. (2010), [17] Scott (1996), [18] Gregory \& Taylor (1986), [19] Bower \& Saul (2011), [20] Croft et al. (2011), [21] Bell et al. (2011).}
\tablenotetext{a}{Smaller fraction of the sky was observed more than twice.}
\tablenotetext{b}{The total surveyed area is about 2870\,deg$^{2}$, but about 460\,deg$^{2}$ was surveyed every day. These parameters are deduced from the Kida et al. (2008) and Matsumura et al. (2009) papers (see \S\ref{Prev}).}
\tablenotetext{c}{Marginal detection ($4.3\sigma$). Ignored in Figure~\ref{Fig:ArealDen_Flux_SurveySummary}.}
\tablenotetext{d}{Mean number of epochs per field - Seven fields were observed on 2732 epochs.}
\tablenotetext{e}{In addition, one transient was found by combining two month worth of data and no transients were found by combining 1\,yr worth of data.}
\tablenotetext{f}{In addition, one transient was found by combining two month worth of data and no transients were found by combining 1\,yr worth of data.}
\tablenotetext{g}{Mean number of epochs per field - Seven fields were observed on 2154 epochs.}
\label{Tab:PreviousSurveys}
\end{deluxetable*}
Here we provide a summary of some of the sky surveys listed in Table~\ref{Tab:PreviousSurveys}.

\subsection{Surveys at frequencies below 2\,GHz}

Carilli et al.~(2003) used a deep, single VLA pointing at
1.4 GHz toward the Lockman hole. They found 
that only a small fraction, $\leq 2$\%, of radio sources above a
flux density limit of 0.1 mJy are highly ($>$50\%) variable on 19 day
and 17 month timescales.  No transients were identified. Frail et al. (1994) imaged a
much larger field at 1.4 GHz toward a $\gamma$-ray burst with the
Dominion Radio Astrophysical Observatory
synthesis telescope, making daily measurements for two weeks and then
on several single epochs for up to three months. No transients were
identified on these timescales, and no sources above a flux density
limit of 3.5\,mJy were seen to vary by more than $4\sigma$.

There have also been a number of wide field surveys of the sky
at 1.4 GHz.
de Vries et al. (2004) used the VLA to image a region, toward
the South Galactic cap, twice on a seven year timescale. No transients were found
above a limit of 2\,mJy.
Croft et al. (2010; 2011) presented results from
the Allen
Telescope Array Twenty-centimeter Survey (ATATS).
They surveyed 690\,deg$^{2}$
of an extragalactic field on 12 epochs.
They compared the individual images with
their combined image and the combined image with the
NRAO VLA Sky Survey (NVSS; Condon et al. 1998).
No transients were found above a
flux density limit of 40 mJy in the combined image, with respect to the NVSS survey (Croft et al. 2010).
In addition, no transients were found in the
individual epochs above flux density of about 100\,mJy (Croft et al. 2011).

A systematic search for transients was
made between the two largest radio sky surveys,
Faint Images of the Radio Sky at Twenty-Centimeters
(FIRST; Becker et al. 1995) and
NVSS (Condon et al. 1998) by Levinson et al. (2002).
Nine transient candidates were identified.  Follow-up observations of
these established that only one was a genuine transient -- a likely
radio supernova in NGC\,4216
(Gal-Yam et al. 2006; Ofek et al. 2010).
We note that each FIRST and NVSS image is composed of $\approx4$ overlapping beams
of adjacent regions taken typically with $\delta{t}\sim$days
(Becker et al. 1995; Condon et al. 1998; Ofek \& Frail 2011).
Therefore, if the duration ($t_{{\rm dur}}$) of the Bower et al.
transients is shorter than this typical time between images
composing ``one epoch'' then the sensitivity of these surveys for transients is degraded by
$\approx\sqrt{4}$.
However, the Levinson et al. survey used a flux density limit of 6\,mJy
which is ($\gtorder2$) higher than the flux limit of these surveys.
Therefore, its efficiency for Bower et al. like
transients is not degraded.

Bright ($>1$\,Jy), short-lived transients,
have been reported by the
Nasu 1.4 GHz survey (e.g., Matsumura et al. 2009; see \S\ref{sec:Introduction}).
Croft et al. (2010; 2011) argued that these transients are not real
because their implied event rate cannot be reconciled with their own survey
unless this population has
sharp cut off at flux densities below 1 Jy.
We note that Croft et al. adopted the Nasu transients
areal density
reported in Matsumura et al. (2009).
However, this areal density is inconsistent with the rate reported
in Kida et al. (2008) which is roughly two orders of magnitude lower.
Furthermore, based on the Nasu survey parameters reported
in Matsumura et al. (2009) we estimate that the areal density
of the Nasu transients is roughly two orders of magnitude lower
than that stated in their
paper\footnote{This inconsistency was already mentioned by
Bower \& Saul (2011). However, they reach a different conclusion than we do.
More information about the Nasu survey is needed in order to resolve this issue.}.

Here we present an estimate of the rate of transients from
the Nasu observations:
Matsumura et al. (2009) reported that they discovered (at that time)
nine transients over a period of two years (730\,days).
Assuming that they have
four pairs of antennas ($N_{ant}$) each looking at a different position
(near the local zenith $\delta\approx 37$\,deg)
with a field of view of $W=0.4$\,deg and scanning the sky at the
sidereal rate, the total
sky area scanned by their system after two years
is $\approx 3.4\times10^{5}$\,deg$^{2}$ ($\cong N_{ant}W \times 360\cos(37^{\circ})\times730$).
Therefore,
the total areal density of their transients is 
the number of transients divided by the total scanned sky area
and divided by 1.61, which is
$\approx1.7\times10^{-5}$\,deg$^{-2}$.
Where the 1.61 factor is due to the fact that their sensitivity is not
uniform within the beam and, and is calculated using Equation~\ref{Eq:NbUni} in
Appendix~\ref{Ap:Rate}.
Assuming the transients duration is 1\,day than their rate is
$\approx 0.02$\,deg$^{-2}$\,yr$^{-1}$.
We note that our derived Nasu transients rate
is roughly consistent with the upper limit on the rate
given by Kida et al. (2008).
Therefore,
we speculate that there was a confusion between areal density and transient
rate in the Matsumura et al. (2009) paper.
We conclude that if our estimate for the areal density
for the Nasu transients is correct than
the ATATS survey results cannot decisively rule out the reality
of the Nasu transients.

A comprehensive survey at 0.8\,GHz was reported
by Bannister et al. (2010).
They surveyed 2776\,deg$^{2}$ south of $\delta=-30^\circ$ over a 22-year period.
Out of about 30,000 sources they identified 53 variables and 15 transient
sources.
Recently, Bower \& Saul (2011) reported a transient search in
the fields of the VLA calibrators in which no transients were found
(Summarized in Table~\ref{Tab:PreviousSurveys} and Figure~\ref{Fig:ArealDen_Flux_SurveySummary}).
Another related work by Bell et al. (2011) searched for radio
transients in the fields of the VLA phase calibrators at 1.4\,GHz, 4.9\,GHz and 8.5\,GHz.
Based on their survey parameters (Table~\ref{Tab:PreviousSurveys})
we estimate that their $95\%$ confidence surface density upper limit on transients
brighter than 8\,mJy 
in 1.4\,GHz, 4.9\,GHz and 8.5\,GHz
are 0.19\,deg$^{-2}$, 0.13\,deg$^{-2}$ and 0.50\,deg$^{-2}$, respectively.
These values are corrected for the beam non-uniformity factor
(of 1.61), mentioned earlier.
We assumed that Bell et al. (2011) searched for transients within the full width at half power
of the beam.
For clarity purposes,
in Figure~\ref{Fig:ArealDen_Flux_SurveySummary} we show only the 4.9\,GHz limit.

\subsection{Surveys at frequencies above 2\,GHz}

There have been several additional transient surveys
carried out at frequencies
above 1.4 GHz.
At 3.1 GHz Bower et al.
(2010) report a marginal detection of one possible transient ($4.3\sigma$) in a
10\,deg$^2$ survey of the Bo\"otes extragalactic field. In a five year
catalog of radio afterglow observations of 75 $\gamma$-ray bursts,
Frail et al. (2003) found several strong variables at 5 and 8.5
GHz, but no new transients apart from the radio afterglows themselves.

Two surveys at 5\,GHz have specifically targeted the Galactic plane.
Taylor and Gregory (1983) and Gregory and Taylor
(1986) used the NRAO 91-m telescope to image an
approximately 500\,deg$^2$ region from Galactic longitude $l=40$\,deg to $l=220$\,deg
with Galactic latitude $\vert{b}\vert\leq 2$\,deg in 16 epochs over a 5-year period.
They identified one transient candidate which underwent a 1\,Jy flare
but for which follow-up VLA observations showed no quiescent radio
counterpart (Tsutsumi et al.~1995). They also claimed tentative evidence for a
separate Galactic population of strong variables comprising 2\% of their
sources. Support for this comes from Galactic survey of Becker et al.
(2010), who find about one half of their variable source
sample (17/39), or 3\% of all radio sources in the Galactic plane, undergo
strong variability on 1-year and 15-year baselines.
We note that the surface density of radio sources in
the Galactic plane is only slightly higher ($\approx 20$\%)
than at high Galactic
latitudes (Helfand et al. 2006; Murphy et al. 2007).

One of the largest variability survey of its kind was carried out using the
7-beam receiver on the NRAO 91-m telescope (Scott 1996; Gregory et al. 2001).
The sky from 0$^\circ \leq \delta\leq 75^\circ$
was surveyed over two 1-month periods in 1986 November and 1987 October
(Condon, Broderick \& Seielstad 1989;
Becker et al. 1991;
Condon et al. 1994;
Gregory et al. 1996).
The final catalog, made by combining both the 1996 and 1997 epochs,
contained 75,162 discrete sources with flux densities $>18$ mJy. 
Long term variability information was available for the majority
of the sources
by comparing the mean flux densities between the 1986 and 1987 epochs.
Scott (1996) carried out a preliminary analysis of the long-term
measurements and identified 146 highly variable sources, or $<$1\% of
the cataloged radio sources.

Eight possible transients
in the Scott (1996; Table 5.1) list appear in either 1986 or 1987
but are undetected in the other epoch ($<2$$\sigma$). Two sources are previously identified 
variables from the Gregory and Taylor (1986) survey, while six are flagged as 
possible false positives due to confusion by nearby bright sources. One source
(B150958.3$+$103541) was $9\pm6$\,mJy in 1986 and $75\pm7$\,mJy in 1987
but it is in both the FIRST and NVSS source catalogs. There are therefore no
long-term transients identified in the Scott (1996) survey.

In order to estimate the flux limit above which the Scott (1996)
comparison between the 1986 and 1987 surveys
is complete, we compared the source numbers near the celestial equator,
as a function of flux in the two publicly available catalogs
from 1987 and the combined 1986/1987 catalog.
We found that at flux densities below about 40\,mJy
the number of sources, as a function of flux,
in the 1987 catalog is 
rising slower than that for the one of the deeper combined catalog.
Therefore, we estimate that near the terrestrial equator,
the 1987 catalog is complete
above a flux density of about 40\,mJy.
However, these catalogs were made from observations
in which each point on the sky was observed
$\approx 4/\cos(\delta)$ times taken within a few days.
This degrades the sensitivity of the comparison carried out by Scott (1996),
for $\ltorder 1$\,day transients,
by about $\sqrt{4}$.
Therefore, we conclude that the Scott (1996) survey is sensitive to short term ($\ltorder 1$\,day)
transients brighter than about 80\,mJy ($=40\sqrt{4}$).
Finally, assuming Scott (1996) did~not find any transients in two epochs,
we put a $2$-$\sigma$ upper limit on the areal density
of $\ltorder 1$\,day transients
brighter than 80\,mJy, of $9.5\times10^{-5}$\,deg$^{-2}$.

In summary, despite the heterogeneous  nature of  these GHz surveys, it 
is clear that the radio sky is relatively quiet compared
with the $\gamma$-rays sky.
The fraction of strong variables
among the persistent radio source population is $0.1$--$3$\% from flux densities of  0.1\,mJy
to 1\,Jy. 
However, the exact percentage of strong variables is still uncertain because of the 
different criteria used by various surveys.
We note that
within this flux density range, radio source populations are dominated by AGN 
roughly above 1\,mJy and star forming galaxies dominates the source counts
below 1\,mJy (Condon 1984; Windhorst 1985).

\begin{figure*}
\centerline{\includegraphics[width=16cm]{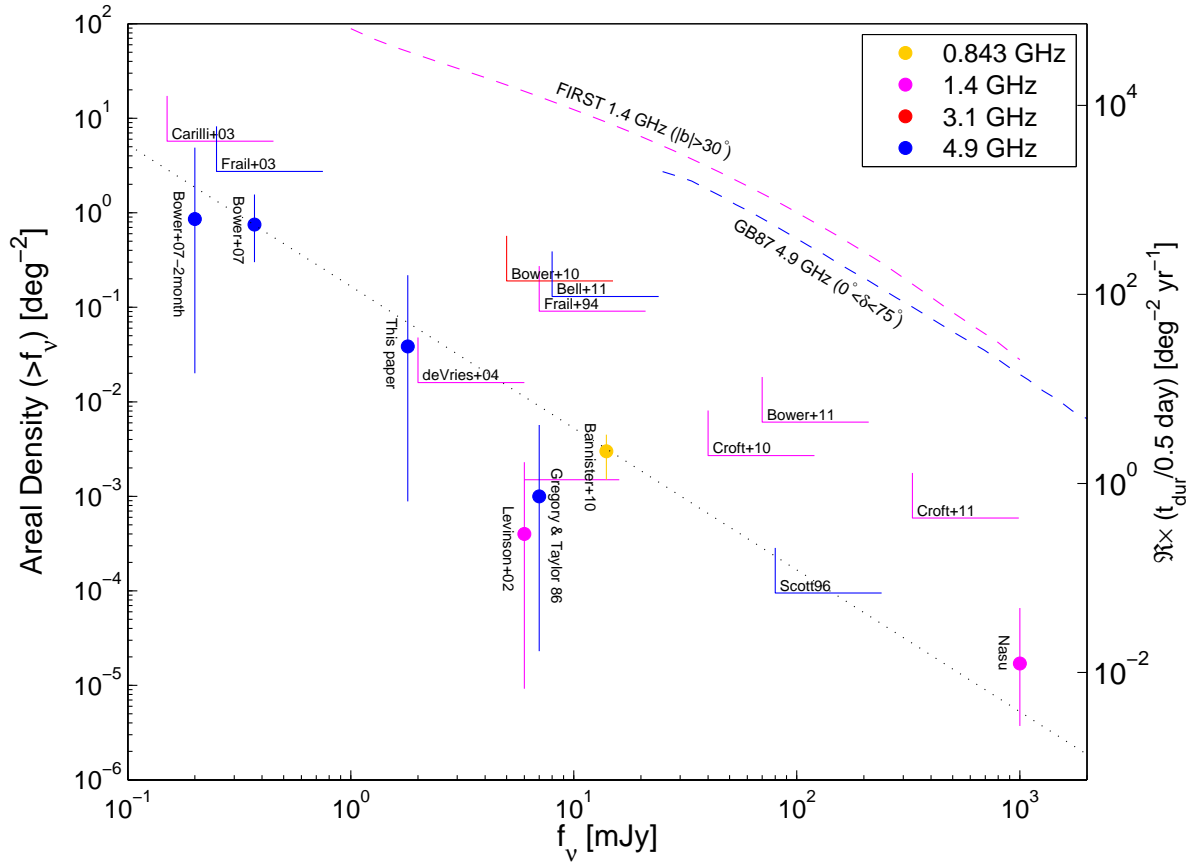}}
\caption{Cumulative areal density of radio sources and transients as a function
of flux density for various surveys.
Different colors represent different frequencies as specified in the legend.
95\% confidence upper limits from various transient surveys are shown
as right-angle corners, while measured areal densities are marked as filled circles.
All the error bars represent 2-$\sigma$ confidence intervals.
We note that the Nasu survey rate is based on our estimate using the
survey description in their papers (see \S\ref{Prev}).
Because some of the parameters
of this survey are unknown to us, we increase the error bars for
this survey to include a factor of two uncertainty.
For the Levinson et al. survey we mark the
areal density of the single transient found in this
survey (a supernova in NGC~4216)
and also the 95\% confidence upper limit
assuming there are no Bower et al. transients in the survey.
For the de Vries et al. (2004) search we use a flux limit of 2\,mJy
since they used the FIRST survey in which each epoch is
composed of about four observations of the same field taken within $\delta{t}\approx$days.
Therefore, this degrade their survey flux limit sensitivity to Bower et al. transients by a factor
of about $\sqrt{4}$.
The right-hand-side y axis shows the transients rate
assuming a transient duration of 0.5\,days.
Some surveys are excluded from this plot.
For example,
Becker et al. (2010) restricted their catalog to sources
detected in at least two out of three epochs,
or that have a confirmed detection at 1.4\,GHz.
Therefore, such surveys are not included here.
Also shown are the areal densities of persistent sources at 1.4\,GHz
and 4.9\,GHz based on the FIRST and GB87 surveys,
respectively (dashed lines).
\label{Fig:ArealDen_Flux_SurveySummary}}
\end{figure*}

The transient areal densities detected by these various surveys,
as well as our survey,
are shown graphically in Figure~\ref{Fig:ArealDen_Flux_SurveySummary}.
Also shown in this figure are the persistent sources
areal densities at different frequencies.
This plot is further discussed in \S\ref{Disc}.

\section{Survey Observations}
\label{SurveyObs}

We designed a survey to look for transients and variable sources near
the Galactic plane, with typical time scales of days to two years
at milli-Jansky flux levels.  We were specially interested in finding
``Bower et al. transients'', conduct multi-wavelength follow up of these
events and finding counterparts and studying their spectral
evolution.

\subsection{Survey Design}
\label{Design}

We used the VLA to observe 141 pointings along the Galactic
plane.  In order to minimize telescope motions we selected all the
pointings in four regions.  The median longitude ($l$) and latitude
($b$) of the four regions are: $l=22.6$\,deg, $b=-6.7$\,deg;
$l=56.6$\,deg, $b=-5.5$\,deg;
$l=89.7$\,deg, $b=-7.8$\,deg; $l=106.0$\,deg, $b=-6.5$\,deg.
Within each region we
selected 26--42 pointings within 2.3\,deg from the median position of
each region.  Each pointing was selected to have no NVSS sources
brighter than 1\,Jy within 3\,deg, no NVSS sources brighter than
300\,mJy within 1\,deg, and no source brighter than 100\,mJy within
the field of view as defined by the half power radius.
We also rejected fields which distance from known Galactic supernova remnants (SNR; Green 2001)
is within twice the diameter of the SNR.
The typical distance between pointings in each
region is about $20'$.  The final 141 pointings are listed in
Table~\ref{Tab:ListOfFields}.
\begin{deluxetable}{llll}
\tablecolumns{4}
\tabletypesize{\scriptsize}
\tablewidth{0pt}
\tablecaption{List of survey pointings}
\tablehead{
\colhead{Field Name}   &
\colhead{RA}   &
\colhead{Dec}  &
\colhead{$N_{{\rm ep}}$} \\
\colhead{}   &
\colhead{deg}   &
\colhead{deg}   &  
\colhead{}   
}
\startdata
1851$-$1327 & $282.94699$ & $-13.45500$ & 16\\
1852$-$1309 & $283.09963$ & $-13.15668$ & 16\\
1853$-$1233 & $283.40439$ & $-12.56005$ & 16\\
1853$-$1251 & $283.25209$ & $-12.85836$ & 16\\
1853$-$1318 & $283.40625$ & $-13.30508$ & 16
\enddata
\tablecomments{List of all 141 fields that were observed as part of this survey. The number of epochs per pointing is marked in $N_{{\rm ep}}$. This table is published in its entirety in the electronic edition of the {\it Astrophysical Journal}. A portion of the full table is shown here for guidance regarding its form and content.}
\label{Tab:ListOfFields}
\end{deluxetable}

\subsection{Observations}
\label{Obs}

These 141 fields were observed on 11 epochs using the VLA
in the Summer of 2008 and on five epochs using the Expanded VLA (EVLA)
during the Summer of 2010.
All observations were made in the compact D configuration.
For the 2008
observations, we added together two adjacent 50 MHz bandwidths
centered at 4835 and 4885\,MHz with full polarization.
For the 2010 observations we added together two adjacent 128\,MHz
sub bands centered at 4896 and 5024\,MHz with full polarization.

In 2008 care was taken to ensure that the local sidereal start time
was the same for each 3-hr epoch (20:30 LST).
Therefore, each field
was observed at the same hour angle and subsequently the synthesized
beam stayed the same for each epoch, varying only when antennas were
taken out of the array. The 2010 observations were taken during EVLA
shared-risk science commissioning, and so some scans lost due to
correlator errors and the last two epochs began one hour earlier than
our 2008 local sidereal start time.

We integrated each pointing for about 50\,s on average.  The maximum
integration time was 58.5\,s and the minimum was 43.3\,s. Additionally,
during each 3-hr observing run we carried out all necessary
calibrations.  Amplitude calibration was achieved with observations of
3C\,286 and 3C\,147 at the start and end of each epoch, respectively.
Phase calibration was checked every 20-25\,min by switching to a bright
point source within a few degrees of the targeted region.
We used the following four phase calibrators (one per each region):
J1911$-$201, J1925$+$211, J2202$+$422, and J2343$+$538.
The total
calibration and
antenna move-time overhead was about $30$\% of the observing time.
This overhead on move time
could have been lowered had we used the fast slew methods from the
NVSS and FIRST surveys (Condon et al.  1998, Becker et al. 1995) with
a resulting increase in the number of square degrees of sky surveyed
per hour.  However, since we recently found that this method could
introduce spurious transients (Ofek et al. 2010), we adopted a less
efficient but more robust observing method.

In Table~\ref{Tab:ListOfEpochs} we list the time of the UTC midpoint
\begin{deluxetable*}{lllccccc}
\tablecolumns{8}
\tablecaption{Observing Epochs}
\tablehead{
\colhead{Epoch} &
\colhead{Date} & 
\colhead{Time Elapsed} &
\colhead{$\langle$rms Noise$\rangle$}&
\colhead{Observed} &
\colhead{Number} &
\colhead{Gain Corr.} &
\colhead{Cosmic error} \\
\colhead{}   &
\colhead{UTC} & 
\colhead{days} & 
\colhead{$\mu$Jy} &
\colhead{Fields} &
\colhead{of Sources} &
\colhead{} & 
\colhead{$\%$} 
}
\startdata
1 & 2008 Jul 15.40 & 0.00    &  243 & 141 & 343 & 1.029 & 5.3 \\
2 & 2008 Jul 18.39 & 2.99    &  181 & 141 & 155 & 0.987 & 4.6 \\
3 & 2008 Jul 19.39 & 3.99    &  229 & 141 & 363 & 1.033 & 4.6 \\
4 & 2008 Aug 10.33 & 25.93   &  178 & 141 & 166 & 0.971 & 1.0 \\
5 & 2008 Aug 11.33 & 26.93   &  183 & 141 & 164 & 0.980 & 1.1 \\
6 & 2008 Aug 14.32 & 29.92   &  174 & 141 & 162 & 1.002 & 0.4 \\
7 & 2008 Aug 16.31 & 31.91   &  173 & 139 & 151 & 1.023 & 1.7 \\
8 & 2008 Aug 18.29 & 33.89   &  178 & 141 & 155 & 0.924 & 1.7 \\
9 & 2008 Aug 25.29 & 40.89   &  196 & 141 & 183 & 1.016 & 2.4 \\
10& 2008 Aug 28.28 & 43.88   &  178 & 141 & 157 & 0.956 & 3.5 \\
11& 2008 Aug 30.28 & 45.88   &  186 & 141 & 170 & 1.033 & 5.3 \\
12& 2010 Jul 16.42 & 731.02  &  105 & 141 & 216 & 1.020 & 1.9 \\
13& 2010 Jul 18.43 & 733.03  &  111 & 134 & 199 & 1.022 & 0.8 \\
14& 2010 Jul 22.35 & 736.95  &  108 & 109 & 158 & 0.992 & 0.6 \\
15& 2010 Jul 23.32 & 737.92  &  104 & 141 & 200 & 1.008 & 0.7 \\
16& 2010 Jul 25.31 & 739.91  &  116 & 140 & 217 & 1.003 & 2.1
\enddata
\tablecomments{List of the 16 epochs. The dates indicate the
observations mid-time. In practice, in all the instances in which we find
that the cosmic error is smaller than 3\% we replaced it by 3\% (see \S\ref{PostSurveyCat}).}
\label{Tab:ListOfEpochs}
\end{deluxetable*}
for each epoch with some additional information.  The shortest
variability timescale sampled was 24\,hr and the longest was
$\cong2$\,yr. By design, the cadence of the 2008 survey was chosen to
probe variability timescales between
a day and a month. A longer 2\,yr timescale was also
sampled by comparing deep images made from the 2008 and 2010 campaigns
(see below).

\section{Data reduction and calibration}
\label{Red}

In 2008 the {\it uv} data was streamed directly to a disk in real time,
and a pipeline was ran after each one of the four regions were
observed.  We used the data reduction pipeline provided in the
Astronomical Image Processing System (AIPS)
package\footnote{http://www.aips.nrao.edu/}.
For each epoch, the pipeline first flagged and calibrated the {\it uv}
data. It then imaged a 30-arcmin-wide field around all 141 pointings,
deconvolving down to three times the rms noise, and restoring the image
with a robust weighted beam.  No self-calibration was done. The VLA
data rates ($\approx 30$\,Mbytes\,hr$^{-1}$) and the D-configuration image
requirements (512 pixels, 3.6$^{\prime\prime}$\,pixel$^{-1}$) were so modest
that the entire pipeline reduction and the variable source analysis
(see \S\ref{RealTimeTran}) was completed before the VLA finished observing the
next region ($\approx$40 min). The real-time analysis
capability was not available in 2010 but the data were also calibrated
within AIPS following standard practice.

As the experiment progressed we built up reference images, made by
summing all previous epochs. These deeper images proved useful in the
real-time search for transient sources (\S\ref{RealTimeTran}). After the
survey was completed, a final set of images was made separately for
each yearly campaign using the data from the 11 epochs in 2008
and the five epochs in 2010. We also summed the 2008 and
2010 deep images to create 16-epoch Master images for the entire
experiment. In summary, there were three final image datasets; the
Single epoch images, the Yearly images (for 2008 and 2010 separately)
and the final Master images made from all available data.

Our final survey parameters are given in Table~\ref{Tab:SurveyPar}.
The effective survey area was calculated using the full width at
half-power ($9.3'$ at 4.86 GHz; see however Appendix~\ref{Ap:Rate})
but the searches for transients in real time and for
variability were made over a larger area -- out to the 15\% response
point of the primary beam ($15'$ diameter).
For the analysis, no correction was made
for the primary beam attenuation, in order to maintain uniform noise
statistics over the entire field.
The synthesized beam
and rms noise estimates for different epochs and different pointings
varied by factors close to unity.  The values in
Table~\ref{Tab:ListOfEpochs} are averages for each epoch over all
pointings, while Table~\ref{Tab:SurveyPar}
gives the mean rms values (over all fields) in the Master images and the
2008 and 2010 combined images.
\begin{deluxetable}{ll}
\tablecaption{Survey Parameters}
\tablehead{
\colhead{Property} &
\colhead{Value}\\
}
\startdata
Frequency                     &       4.9 GHz \\
Observing time                &       48 hrs \\
Survey Area                   &       2.66 deg$^2$ \\
Angular resolution            &       15$^{\prime\prime}$ \\
Repeats                       &       16 \\
Timescales                    &       1\,day--2\,years\\
No. of fields                 &       141 \\
mean exposure Time per field  &       50 s \\
mean rms per epoch (2008)     &       190 $\mu$Jy \\
mean rms per epoch (2010)     &       109 $\mu$Jy \\
mean rms per 11 epochs (2008) &       72 $\mu$Jy \\
mean rms per 5 epochs (2010)  &       56 $\mu$Jy  \\
mean rms in Master images     &       46.7$\mu$Jy 
\enddata
\tablecomments{}
\label{Tab:SurveyPar}
\end{deluxetable}
Throughout the paper we state explicitly if we use corrected
or uncorrected fluxes.
``Corrected fluxes'' are corrected
for beam attenuation and for the CLEAN bias
by adding additional $+0.3$\,mJy
(e.g., Becker et al. 1995; Condon et al. 1998).
In order to maintain uniform statistics we chose to search all images
to the same depth, rather than compute a new threshold for each image
individually. In practice, this led to some false positives for
noisier than average epochs and fields with bright point sources.
Indeed, Table~\ref{Tab:ListOfEpochs} indicates that the nosiest
epochs contains larger number of sources.

\section{Real-Time Transient Search}
\label{RealTimeTran}

We employed two distinct analysis strategies for transient and
variable source identification. The first, which is discussed in this section,
was a real-time analysis.
The main motivation was to rapidly identify any short-lived
sources and mark them for immediate follow-up
at other wavelengths. The second was a post-survey analysis which was
carried out after all the epochs had been observed.
The main goals of
this second phase were to carry out a more in-depth search for
Bower et al. transients (\S\ref{PostSurveyCat})
and to characterize the variability properties
of the persistent source population (\S\ref{PostSurveyTran}).

For the real-time identification
the images (\S\ref{Red}) were searched visually for any new or
strongly varying sources by comparing them with individual epochs,
and by comparing them with a reference image
made by summing all previous epochs.
Any candidate
variable source which we identified was subject to a more detailed light
curve and position fitting analysis before deciding to trigger radio, optical
and/or X-ray follow-up observations.

The followup visible light observations were carried out using the
robotic Palomar $60''$ telescope (P60; Cenko et al. 2006)
and the Keck-I~10-m telescope. The UV and X-ray observations
were conducted by the {\em Swift} satellite (Gehrels et al. 2004).
We note that prior to and during the VLA campaign we obtained
visible light reference images for most of our fields
using the P60 telescope.

We identified two possible transients that were deemed interesting
enough for multi-wavelength follow-up.
However, followup VLA observations and a careful post observing
re-analysis (\S\ref{PostSurveyTran})
showed that these are not real transients.
One source, J213438.01$+$414836.0,
was a sidelobe artifact,
while the second source, J230424.68$+$530414.7
is a long term variable that had crossed our single-epoch noise
threshold on 2008 July 19 and is clearly seen in the 2008 and 2010 deep
coadds.
We note that both sources were observed using
the P60 telescope about 2\,hr and 1\,hr
after the radio observations were obtained, respectively.
Furthermore {\it Swift}-XRT observations of the
first source were obtained about five days after it was found.
These fast response observations
demonstrate our near real time followup capabilities.

\section{Post survey source catalog}
\label{PostSurveyCat}

The next phase of our analysis occurred after
the conclusion of the observations.
We generated source catalogs in order to search for
short-lived transients and to carry out a variability study of all
identified sources. The AIPS task {\it SAD} (search and destroy) was used for source
finding.

We found that
false positive sources came from one of two main reasons: slightly resolved
sources and sidelobe contamination.
Extended sources were
identified by requiring that their integrated flux density was within a
factor of two of their peak flux density.
False sources created by
scattered power from the snapshot sidelobe response was only a
significant problem for the six fields with sources whose flux density
exceeded 20\,mJy. We flagged any variables or transients from these
fields for visual inspection.

Three catalogs were created. The first catalog is the ``Single epoch
catalog'', generated by running {\it SAD} on a 15-arcmin diameter
region for each single-epoch
image individually (Table~\ref{Tab:SingleEpochCat}).
A second ``Master
catalog'' (Table~\ref{Tab:MasterCat})
was created by running {\it SAD} on the final master
images made from all available data (16 epochs), while a third catalog,
``Yearly catalog'', was generated on the yearly images (for 2008 and
2010 separately).
\begin{deluxetable*}{lllllllllllll}
\tablecolumns{13}
\tabletypesize{\scriptsize}
\tablewidth{0pt}
\tablecaption{Single epoch catalog}
\tablehead{
\colhead{Epoch}                        &
\colhead{Source}                       &
\colhead{Field Name}                   &
\colhead{J2000 RA}                     &
\colhead{J2000 Dec}                    &
\colhead{$f_{\nu,{\rm p}}^{{\rm cor}}$}   &
\colhead{$f_{\nu,{\rm p}}^{{\rm uncor}}$} &
\colhead{$\sigma_{{\rm p}}$}            &
\colhead{$f/f_{{\rm p}}$}               &
\colhead{Major}                       &
\colhead{Minor}                       &
\colhead{PA}                          &
\colhead{$\Delta_{{\rm C}}$}           \\
\colhead{}                            &
\colhead{}                            &
\colhead{}                            &
\colhead{deg}                         &
\colhead{deg}                         &
\colhead{mJy}                         &
\colhead{mJy}                         &
\colhead{mJy}                         &
\colhead{}                            &
\colhead{$''$}                        &
\colhead{$''$}                        &
\colhead{deg}                         &
\colhead{$'$}           
}
\startdata
 1 &    1 &   $1851-1327$ &   282.955970&  $-13.502335$&  329.63&  25.39 &  0.26 &  1.03 &   23.67 &  13.67 &  23.7 &    2.89\\
 1 &    2 &   $1851-1327$ &   282.937502&  $-13.432227$&    3.29&   1.23 &  0.26 &  1.00 &   23.49 &  13.31 &  29.7 &    1.48\\
 1 &    3 &   $1851-1327$ &   282.855004&  $-13.349517$&    8.60&   1.12 &  0.26 &  0.64 &   23.66 &   8.48 &  13.6 &    8.30\\
 1 &    4 &   $1853-1233$ &   283.399554&  $-12.561197$&    7.59&   6.59 &  0.26 &  0.95 &   21.88 &  13.15 &  20.3 &    0.29\\
 1 &    5 &   $1853-1233$ &   283.474902&  $-12.562245$&    4.74&   3.10 &  0.26 &  0.89 &   22.97 &  11.74 &  21.5 &    4.13
\enddata
\tablecomments{Catalog of 2953 sources detected in Single epochs.
{\it Columns description:} Epoch is the epoch number (see Table~\ref{Tab:ListOfEpochs}), and source is a serial source index in epoch.
$f_{\nu,{\rm p}}^{{\rm cor}}$, $f_{\nu,{\rm p}}^{{\rm uncor}}$, and $\sigma_{{\rm p}}$
are the corrected peak flux density, uncorrected peak flux density,
and the error in the uncorrected peak flux density, respectively.
Corrected flux are corrected for beam attenuation and the
CLEAN bias.
$f/f_{{\rm p}}$ is the integrated flux divided by the peak flux.
Major and Minor are the major and minor axes object size, while PA is the position angle
of the major axis.
Finally, $\Delta_{{\rm C}}$ is the distance of the source from the beam center.
This table is published in its entirety in the electronic edition of the {\it Astrophysical Journal}.
A portion of the full table is shown here for guidance regarding its form and content.
\label{Tab:SingleEpochCat}}
\end{deluxetable*}

The Master catalog has the merit of being able to identify persistent
radio sources approximately $\sqrt{N}$-times fainter than any
individual epoch (where $N=16$), but it is $\sqrt{N}$-times less
sensitive to a short-lived transient that might be identified in a single-epoch image.
For our Single epoch catalogs we used 
a flux density cutoff of 1\,mJy in 2008 and 0.76\,mJy in 2010,
and the number of sources in each epoch are given in Table~\ref{Tab:ListOfEpochs}.
Our Master catalog consisted of 464 sources which
are listed in Table~\ref{Tab:MasterCat}, with a flux density
cutoff of $0.28$\,mJy.  The Yearly catalog had flux density cutoff of
0.5\,mJy in 2008 and $0.35$\,mJy in 2010. These cutoffs corresponds to
about a 5-7$\sigma$ threshold, depending on the rms noise for
individual fields.
The electronic version of the Master catalog also contains
the peak flux of each source.
This was measured in
the Single epoch images
at the position of the sources
found in the Master image.

We used the Master catalog to perform a second order amplitude
calibration that would tie together the flux density scale for all
epochs. Normally self-calibration could be used to find additional
gain variations within a radio observation but our survey was designed to
avoid pointings with bright point sources.
Our approach assumes that each VLA epoch (all the observations
in each epoch were taken within 3\,hours) shares the same ``gain''
correction, and we solved for these ``nightly'' gain corrections by
fitting, using least squares minimization, the equation
\begin{equation}
m_{ij} = z_{i} + \bar{m}_{j},
\label{Eq:RelPhot}
\end{equation}
where $m_{ij}$ is the ``magnitude'': $-2.5\log_{10}{f_{ij}}$, $f_{ij}$
is the peak specific flux of the $j$-th source in the $i$-th epoch, $z_{i}$ is
the gain correction for the $i$-th epoch (in units of magnitudes)
and $\bar{m}_{j}$ is a
nuisance parameter representing the best fit mean magnitude of the
$j$-th source.
We note that, as explained in Appendix~\ref{Ap:ZP},
magnitudes have convenient statistical properties.
The final multiplicative gain corrections are
$10^{-0.4z_{i}}$.
This method is described in detail in
Appendix~\ref{Ap:ZP}, and the best fit multiplicative gain corrections are
listed in Table~\ref{Tab:ListOfEpochs}.
Similarly, we also derived the yearly gain corrections for the
Yearly epochs.
These gain
corrections are $1.004$ and $0.996$ for 2008 and 2010, respectively.

The flux errors reported by {\it SAD} do~not include
any systematic error terms.
Typically, VLA calibration is assumed to be good
to a level of $3\%$ or better (e.g., Condon et al. 1998). 
In order to check if some epochs are nosier we estimated
the ``cosmic errors'', $\epsilon_{{\rm cos}}$, using the following scheme.
We measured
the standard deviation in the flux of the four phase calibrators
observed on each night, after normalizing their flux by their mean
flux over all the epochs taken at the same year.  The cosmic errors
estimated using this method are listed for each epoch in
Table~\ref{Tab:ListOfEpochs}.
This estimate is based on a small number of sources
and these sources may be variable.
Therefore, this should be regarded as a rough estimate.
In some instances, the cosmic errors we estimated
were smaller than $3\%$ and in those cases we replaced
the cosmic errors for these epochs by $3\%$.
We note that if indeed the 
cosmic errors in some cases are smaller than 3\%,
then our strategy of adopting a larger cosmic errors
may reduce the number of variables
found in our survey.
For the Yearly catalogs, we used the
mean cosmic error terms of the individual epochs in each year.
These are $0.028$ and $0.012$ for 2008 and 2010, respectively.

Equipped with the gain corrections and an estimate for the cosmic errors
we next corrected the flux measurements of all the sources using the
gain correction factors and added in quadrature the cosmic errors to the
peak flux errors.
The new fluxes and errors were used in all the plots
and the calculation of the light curves statistical properties.
\begin{deluxetable*}{rrrrlllrlrllrll}
\tablecolumns{15}
\tabletypesize{\scriptsize}
\tablewidth{0pt}
\tablecaption{Master catalog}
\tablehead{
\multicolumn{11}{c}{Current search} &
\multicolumn{2}{c}{NVSS} &
\multicolumn{1}{c}{USNO} &
\multicolumn{1}{c}{2MASS} \\
\colhead{J2000 RA}   &
\colhead{J2000 Dec}   &
\colhead{$f_{{\rm p}}^{{\rm cor}}$} &
\colhead{$f_{{\rm p}}^{{\rm uncor}}$} &
\colhead{$\sigma_{{\rm p}}$} &
\colhead{$N_{{\rm obs}}$} &
\colhead{$N_{{\rm det}}$} &
\colhead{$\chi^{2}$} &
\colhead{StD/$\langle f\rangle$} &
\colhead{$\chi^{2}_{{\rm Y}}$} &
\colhead{StD/$\langle f\rangle_{Y}$} &
\colhead{Dist} & 
\colhead{$\alpha$\tablenotemark{a}} &
\colhead{Dist} &
\colhead{Dist} \\
\colhead{deg}   &
\colhead{deg}   &
\colhead{mJy} &
\colhead{mJy} &
\colhead{mJy} &
\colhead{} &
\colhead{} &
\colhead{} &
\colhead{} &
\colhead{} &
\colhead{} &
\colhead{$''$} & 
\colhead{} &
\colhead{$''$} &
\colhead{$''$}
}
\startdata
 283.675661&$-13.518385$&  7.24&   2.48& 0.05& 15& 15&   50.23&  0.17&   2.54&  0.06&      1.0&     1.03& \nodata& \nodata\\
 283.682476&$-13.444363$&  2.52&   1.85& 0.05& 15& 15&   52.34&  0.18&   0.05&  0.02&      3.8&     0.17& \nodata& \nodata\\
 284.193460&$-13.325061$& 47.83&  32.01& 0.10& 14& 14&   70.30&  0.08&   0.48&  0.02&      0.9&    -0.07& \nodata&     0.6\\
 284.387195&$-12.213881$& 11.03&   6.74& 0.06& 15& 15&  142.60&  0.14&   2.85&  0.05&      4.4&    -0.54& \nodata& \nodata\\
 283.976980&$-12.165968$& 93.19&  57.18& 0.13& 15& 15&  468.40&  0.18&  68.27&  0.25&      0.9&     0.08& \nodata& \nodata\\
 283.949467&$-11.356239$&  7.65&   4.69& 0.05& 15& 14&   53.20&  0.10&   0.84&  0.03&      2.9&     1.14& \nodata& \nodata\\
 284.592955&$-10.949209$& 13.07&  10.66& 0.06& 15& 15&   48.94&  0.08&   3.98&  0.06&      4.2&     1.26& \nodata& \nodata\\
 284.497521&$-10.459850$&  2.70&   2.11& 0.05& 15& 14&   46.02&  0.14&   1.71&  0.06&     10.0&     0.83& \nodata& \nodata\\
 298.720898&$ 16.755257$& 10.92&   8.39& 0.04& 16& 16&  118.76&  0.10&   5.02&  0.08&      1.0&    -0.80& \nodata& \nodata\\
 299.583716&$ 18.171732$&  6.93&   5.82& 0.04& 16& 16&  143.31&  0.14&  15.03&  0.12&      2.3&    -0.20& \nodata& \nodata\\
 299.271523&$ 18.226378$&  4.81&   3.65& 0.05& 16& 16&   75.27&  0.12&   0.67&  0.04&  \nodata&  \nodata& \nodata& \nodata\\
 299.353219&$ 18.473931$& 41.09&   6.89& 0.05& 16& 16&  149.27&  0.13&  13.66&  0.13&      0.4&     0.14& \nodata& \nodata\\
 300.337088&$ 19.936532$&  7.71&   5.50& 0.04& 16& 16&   55.62&  0.08&   0.12&  0.01&      2.2&     0.49& \nodata& \nodata\\
 300.030673&$ 20.034572$& 23.81&   4.23& 0.04& 16& 16&   70.42&  0.11&   0.69&  0.02&      0.6&     0.39& \nodata& \nodata\\
 324.719526&$ 42.022609$&  4.34&   3.83& 0.05& 16& 16&  154.53&  0.16&  20.39&  0.17&  \nodata&  \nodata& \nodata& \nodata\\
 325.523312&$ 42.023491$& 12.85&   2.83& 0.04& 16& 16&  356.39&  0.20&  30.28&  0.20&      1.0&     0.88& \nodata& \nodata\\
 325.348791&$ 42.085546$&  7.61&   2.23& 0.04& 16& 16&  148.57&  0.15&   8.95&  0.12&      0.3&     1.42& \nodata& \nodata\\
 326.519678&$ 42.141397$& 30.49&   8.09& 0.04& 16& 16&   54.95&  0.06&   2.98&  0.05&      0.5&     1.09& \nodata& \nodata\\
 325.966278&$ 42.884463$&  7.55&   2.34& 0.05& 16& 16&   46.73&  0.15&   0.45&  0.02&  \nodata&  \nodata& \nodata& \nodata\\
 326.391328&$ 43.460876$&  7.24&   4.02& 0.04& 16& 16&   46.97&  0.10&   0.32&  0.01&      1.8&     1.00& \nodata& \nodata\\
 341.141889&$ 51.541980$& 15.60&  12.43& 0.05& 16& 16&  160.78&  0.10&   0.53&  0.03&      1.6&     0.54&     0.3& \nodata\\
 342.654997&$ 52.100799$& 24.51&  13.77& 0.05& 16& 16&  195.40&  0.11&  18.74&  0.13&      1.8&     0.28&     0.6&     0.5\\
 341.820302&$ 52.137525$&  4.11&   2.64& 0.04& 16& 16&   49.55&  0.12&   0.43&  0.03&      1.9&     0.66&     0.4&     1.1\\
 343.612547&$ 52.339910$&  9.04&   7.77& 0.05& 16& 16&  168.64&  0.12&   8.37&  0.10&      1.2&    -0.38& \nodata& \nodata\\
 344.074222&$ 52.431561$&  4.50&   2.92& 0.04& 16& 16&  353.19&  0.28&  35.95&  0.24&  \nodata&  \nodata&     1.5& \nodata\\
 342.837121&$ 52.495055$&  3.85&   3.53& 0.04& 16& 16&  102.16&  0.16&   4.92&  0.07&      5.6&     0.39& \nodata& \nodata\\
 346.017848&$ 52.735617$& 10.19&   2.48& 0.04& 16& 16&  181.08&  0.22&   5.30&  0.10&      4.3&    -1.12& \nodata& \nodata\\
 345.504754&$ 52.912710$&  2.13&   1.69& 0.04& 16& 14&   91.42&  0.25&   1.59&  0.06&  \nodata&  \nodata& \nodata& \nodata\\
 345.245407&$ 53.186497$& 33.34&  24.81& 0.05& 16& 16&   46.53&  0.06&   0.86&  0.02&      0.8&     0.91& \nodata& \nodata\\
 346.421593&$ 53.202481$&  8.26&   3.30& 0.05& 16& 16&   71.60&  0.15&   0.18&  0.01&      6.4&     0.35&     0.1& \nodata\\
\hline
 283.976980&$-12.165968$& 93.19&  57.18& 0.13& 15& 15&  468.40&  0.18&  68.27&  0.25&      0.9&     0.08& \nodata& \nodata\\
 298.658519&$ 18.200598$&  3.28&   0.38& 0.04& 16&  0&   41.38&  0.55&  18.72&  0.68&  \nodata&  \nodata&     0.8&     0.3\\
 299.358645&$ 18.324890$&  3.02&   0.87& 0.05& 16&  5&   26.51&  0.22&  23.36&  0.37&      2.0&     1.03& \nodata& \nodata\\
 299.811372&$ 18.369894$&  1.82&   0.94& 0.04& 16&  6&   62.65&  0.29&  22.02&  0.34&  \nodata&  \nodata& \nodata& \nodata\\
 323.729857&$ 41.317576$&  0.91&   0.55& 0.05& 16&  0&   28.52&  0.34&  20.48&  0.53&  \nodata&  \nodata&     1.6&     2.0\\
 324.109846&$ 41.657383$&  1.55&   0.45& 0.05& 16&  0&   70.45&  0.64&  23.78&  0.67&  \nodata&  \nodata& \nodata& \nodata\\
 324.719526&$ 42.022609$&  4.34&   3.83& 0.05& 16& 16&  154.53&  0.16&  20.39&  0.17&  \nodata&  \nodata& \nodata& \nodata\\
 325.523312&$ 42.023491$& 12.85&   2.83& 0.04& 16& 16&  356.39&  0.20&  30.28&  0.20&      1.0&     0.88& \nodata& \nodata\\
 342.654997&$ 52.100799$& 24.51&  13.77& 0.05& 16& 16&  195.40&  0.11&  18.74&  0.13&      1.8&     0.28&     0.6&     0.5\\
 344.074222&$ 52.431561$&  4.50&   2.92& 0.04& 16& 16&  353.19&  0.28&  35.95&  0.24&  \nodata&  \nodata&     1.5& \nodata
\enddata
\tablenotetext{a}{The spectral power-law slope,
defined by $f_{\nu}\propto \nu^{\alpha}$, as measured from
the NVSS 1.4\,GHz specific flux and our 4.9\,GHz corrected flux density.}
\tablecomments{Catalog of the 464 sources detected in the Master images and their properties.
We note that {\it SAD} detected additional sources which are not listed here.
These sources were identified as noise artifacts by subsequent inspection
of the images and were removed from the catalog.
{\it Columns description:}
$f_{\nu,{\rm p}}^{{\rm cor}}$, $f_{\nu,{\rm p}}^{{\rm uncor}}$, and $\sigma_{{\rm p}}$
are described in Table~\ref{Tab:SingleEpochCat}.
$N_{{\rm obs}}$ is the number of epochs in which the source position was observed,
while $N_{{\rm det}}$ is the number of detections in the Single epoch images.
Subscript ``Y'' in $\chi^{2}$ and StD$/\langle f\rangle$ indicates that these
values are calculated for the Yearly epochs.
Dist is the distance between the radio position and the NVSS, USNO-B1 and 2MASS (Skrutskie et al. 2006)
nearest counterparts.
A portion of the full table, containing the 30 Single epoch variable sources
(\S\ref{VarSingle}) and the ten Yearly variable sources (\S\ref{VarYear}) are shown here for guidance regarding its form and content.
The two variable lists are separated by horizontal line.
This table is published in its entirety in the electronic edition of the {\it Astrophysical Journal}.
The electronic table also contains additional columns as the Field name,
sidelobes flag, integrated flux divided by peak flux, major axis, minor axis, and position angle of the sources,
distance from the beam center,
$V_{{\rm R}}$ and $V_{{\rm F}}$ for the yearly fluxes, the NVSS flux, the $B$ and $R$ magnitudes
of the USNO-B 
counterparts, $J$, $H$ and $K$-magnitudes of the 2MASS counterparts 
and all the 16 peak fluxes and errors measurements in the Single epochs and the peak flux
measurements in the Yearly images.
We note that the distance threshold for USNO-B and 2MASS counterparts was set to $2''$,
and $15''$ for NVSS.
\label{Tab:MasterCat}}
\end{deluxetable*}

\section{Post survey transient search}
\label{PostSurveyTran}

\begin{figure}
\centerline{\includegraphics[width=8.5cm]{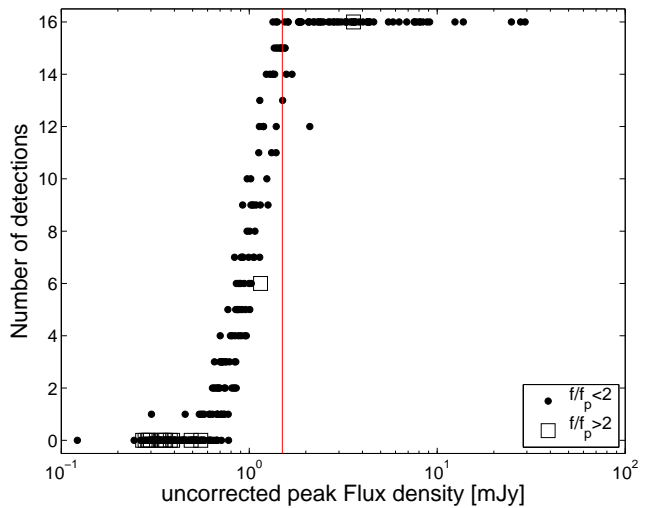}}
\caption{Detection repeatability as a function of flux.
The plot shows, for each source
that its field was observed 16 times
and that was detected in the Master catalog,
the number of {\it SAD} detections in individual epochs.
The number of detections is shown against the uncorrected peak flux
of the source as measured in the Master image.
Point sources are represented by filled circles,
while open squares are for resolved sources.
The vertical line shows the 1.5\,mJy cut we used
to define our completeness limit.
\label{Fig:Flux_Ndet}}
\end{figure}

Our final transient search utilized the catalogs
presented in \S\ref{PostSurveyCat}.
Specifically, we matched all the sources in the individual epochs to
sources in the Master catalog using a $4''$ matching radius.
Sources in individual epochs which do not have counterpart
in the Master catalog are transient candidates.
However, because we used a single flux density threshold,
while the noise varies in each epoch and field,
most of the faint sources are probably noise artifacts.
We used Figure~\ref{Fig:Flux_Ndet} in 
order to choose a reasonable flux density limit for our transient search.
This figure shows the number of detections, in the Single epoch catalogs,
of sources which are detected in the Master catalog and their field
was observed 16 times.
This plot suggests that the probability that a faint source
detected in the master image
will be detected with uncorrected flux density above 1.5\,mJy,
in only one epoch, is low.
In fact it suggests that we could use an even lower threshold.
However, given that this is based on small number statistics
and given the variable quality of different images,
we used an higher flux limit cutoff of 1.5\,mJy.

Following this analysis,
we searched for sources detected in a single epoch
that do~not have a counterpart in the Master catalog,
have uncorrected flux density $>1.5$\,mJy,
and a distance from beam center smaller than $4.65'$
(i.e., half power beam radius).
In total we found 50 sources. However, a close inspection
of these sources shows that most of them are not real.
Of the 50 candidates, 46 are in fields which contain sources brighter
than 10\,mJy, and they are clearly the results of sidelobes.
Of the remaining four candidates,
three sources are also most probably not real.
One candidate was found next to a slightly resolved
5\,mJy source. 
A second object is a known 2\,mJy source for which
the centroid position puzzlingly shifted slightly in one of the epochs,
a third candidate is a sidelobe seen in several epochs,
and the fourth candidate is probably a real transient.
This transient candidate, J$213622.04+415920.3$
is described next.

\subsection{The transient candidate J$213622.04+415920.3$}

We found a single source, J$213622.04+415920.3$,
that was detected only in
the first epoch and may be a real transient.
Given that this source was detected in the first epoch,
before we constructed a reference image,
it was not followed up in real time.
The main properties of this transient candidate are summarized in
Table~\ref{Tab:TranProp}.
The peak flux measurements at the position
of the source at all epochs show that the source
was indeed visible only in the first epoch.
Moreover, this source is not detected
in the Master or Yearly images.
The peak flux at this position
in the Master image is $74\pm42$\,$\mu$Jy,
and in the 2008 (2010) combined images is $161\pm67$\,$\mu$Jy ($-14\pm49$\,$\mu$Jy).
\begin{deluxetable}{ll}
\tablecaption{Properties of the transient candidate J$213622.04+415920.3$}
\tablehead{
\colhead{Property} &
\colhead{Value}\\
}
\startdata
Field name                    &    2136$+$4158         \\
R.A. (J2000.0)                &    $21^{h}36^{m}22.^{s}04 \pm 1.2''$       \\
Dec. (J2000.0)                &    $+41^{\circ}59'20.''3 \pm 1.2''$  \\
Detection date                & 2008 Jul 15.4147         \\
Uncorrected peak flux         &    $1.61\pm0.28$\,mJy    \\
Corrected peak flux           &    $2.36\pm0.41$\,mJy    \\
Distance from beam center     & $2.77'$                
\enddata
\tablecomments{We note that the coordinates in this table are based on Gaussian fit.
The coordinates of this source in the Single epoch Table (Table~\ref{Tab:SingleEpochCat})
 were derived using SAD and they are somewhat different
RA$=21^{h}36^{m}21.^{s}965$, Dec$=+41^{\circ}59'21.''52$ (J2000.0).
Note that the flux is corrected also for the CLEAN bias.}
\label{Tab:TranProp}
\end{deluxetable}
In order to test if the source is variable during the 46\,s VLA integration,
we split the data into two 23-s images. We found that the flux of the source
did~not change significantly between the first and second parts of the exposure.

\begin{figure}
\centerline{\includegraphics[width=8.5cm]{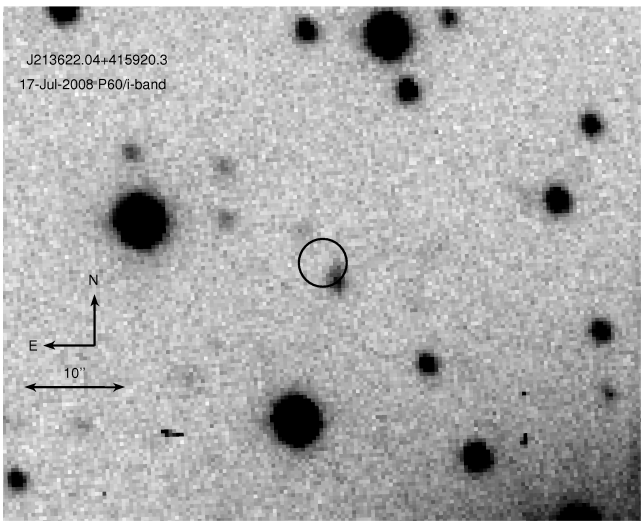}}
\caption{P60 $i$-band image of the field of J$213622.04+415920.3$.
The image, with exposure time of 540\,s, was taken on 2008 Jul 17.45.
A $2.4''$ radius circle marks the transient candidate location.
This radius roughly corresponds to the $2$-$\sigma$ error circle.
\label{Fig:VLA_J213621+415921}}
\end{figure}

The field of J$213622.04+415920.3$ was observed with the P60
telescope in $i$-band about two days after the transient was detected.
This image is shown in Figure~\ref{Fig:VLA_J213621+415921}.
We find two sources near the transient position.
One source is found $\cong3.2''$ North-East of the transient location
and has an $i$-band magnitude of $20.98\pm0.23$.
The other is found $\cong2.3''$
South of the transient position and has a magnitude of $19.46\pm0.20$.
The magnitudes are calibrated relative to USNO-B1.0 I-band
magnitude (Monet et al. 2003).
Given the relatively large angular distances between the transient position
and the nearest visible light sources,
they are probably not associated.
We note that given the stellar density in this region
the probability to find a source within a $2.4''$ radius
from a random position is about $15\%$.
Finally, we do~not find any counterpart to this transient
in the SIMBAD, NED or HEASARC databases.

The search method described in the beginning of this section
may miss transients which are bright enough to be present
in the Master catalog.
Therefore, we also searched for sources which are detected in the Master catalog
and detected in only one of the Single epoch catalogs with uncorrected flux density
above 1.5\,mJy.
No such sources were found.

\section{Variability Analysis}
\label{Variability}

We investigated the variability of all the sources in the Master catalog
using their peak flux densities measured 
in the Single epoch images (\S\ref{VarSingle}),
and the Yearly images (\S\ref{VarYear}).
We used various statistics to assess the
variability and its significance, which we list in Table~\ref{Tab:MasterCat}.
For each
source we calculated its standard deviation (StD) over all the
flux measurements,
the StD over the mean flux (StD/$\langle f\rangle$) and the $\chi^{2}$ given by
\begin{equation}
\chi^{2}=\sum_{i}^{N}{  \frac{(f_{i} - \langle f\rangle )^{2}}{\sigma_{i}^{2}+(f_{i}\epsilon_{{\rm cos},i})^{2}} },
\label{Eq:chi2}
\end{equation}
where $N$ is the number of measurements which is 2 for the Yearly
catalogs and 16 for the
Single epoch catalogs,
$i$ is the epoch index, $f_{i}$ is the gain corrected peak flux in the $i$-th epoch,
$\sigma_{i}$ is its associated error,
$\epsilon_{{\rm cos}}$ is the cosmic error,
and $\langle f\rangle$ is the mean flux of the source over all the epochs.
Note that in Table~\ref{Tab:MasterCat}, $\chi^{2}$ is measured over the individual epochs,
while $\chi^{2}_{Y}$ is measured over the two yearly epochs.
In the first case, the number of degrees of freedom ($dof$) is 15, while in
the second case it is one.

Some previous surveys have defined ``strong variables'' as
exceeding some pre-defined variability measure. There are many
definitions of fractional variability in the literature
and for comparison with other surveys
we list in the electronic version of
Table~\ref{Tab:MasterCat} the two following indicators
\begin{equation}
V_{{\rm R}} = \frac{\max\{f_{i}\}}{\min\{f_{i}\}},
\label{VR}
\end{equation}
and
\begin{equation}
V_{{\rm F}} = \frac{\max\{f_{i}\}-\min\{f_{i}\}}{\max\{f_{i}\}+\min\{f_{i}\}}.
\label{VF}
\end{equation}
We note that these two quantities are related through
$V_{{\rm F}}=(V_{{\rm R}}-1)/(V_{{\rm R}}+1)$.
However, as neither of these indicators account for
measurement errors, 
they cannot be used reliably on their own at low
fluxes where the measurement errors are very large. 
Most importantly, estimators involving the $\min$ or $\max$ functions
strongly depend on the
number of measurements.
This fact complicates direct comparison between different surveys.
For example, a light curve whose fluxes are drawn from
a log normal random distribution, with $N_{{\rm ep}}=3$ survey,
and StD/$\langle f\rangle=0.5$ has $\langle V_{R}\rangle\cong2.44$,
while for $N_{{\rm ep}}=16$ it will have $\langle V_{R}\rangle\cong5.62$.
Therefore, although
Becker et al. (2010) and Taylor and Gregory (1983) 
defined strong variables identically, i.e. $V_{R}\geq $3 ($V_{F}\geq 0.5$),
any direct comparison of these two surveys is difficult since
the number of epochs in these surveys were 3 and 16, respectively.
In the future, in order to cope with this problem we suggest to use the StD/$\langle f\rangle$ as a fractional
variability estimator.
Moreover in Appendix~\ref{Ap:StatVR} we provide a conversion table
for $V_{R}$ as a function of $N_{{\rm ep}}$ and StD/$\langle f\rangle$.

\subsection{Short time scale variability}
\label{VarSingle}

\begin{figure}
\centerline{\includegraphics[width=8.5cm]{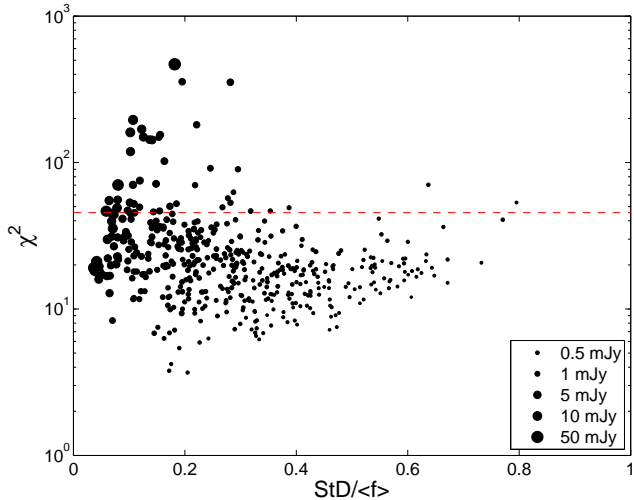}}
\caption{
StD$/\langle f\rangle$ vs. the $\chi^{2}$, where the
uncorrected peak flux of the sources (in the Master catalog)
is marked by symbol size.
The dashed line corresponds to $\chi^{2}>45.5$, which corresponds to $4\sigma$ assuming 15 degrees of freedom.
\label{Fig:VarEp16_StDvsChi2}}
\end{figure}

In order to explore
radio variability on short time scales (e.g., days to weeks),
we constructed 16 epochs light curves for all sources
with peak flux densities larger than 1.5\,mJy ($\approx6\sigma$).
Figure~\ref{Fig:VarEp16_StDvsChi2} shows the
StD$/\langle f\rangle$ vs. the $\chi^{2}$ of all sources in
the Master catalog.
This figure suggests that a large fraction of at least the bright radio sources
with flux densities larger than about 10\,mJy are variables at the level
of $\gtorder5$\%.

\begin{figure*}
\centerline{\includegraphics[width=16cm]{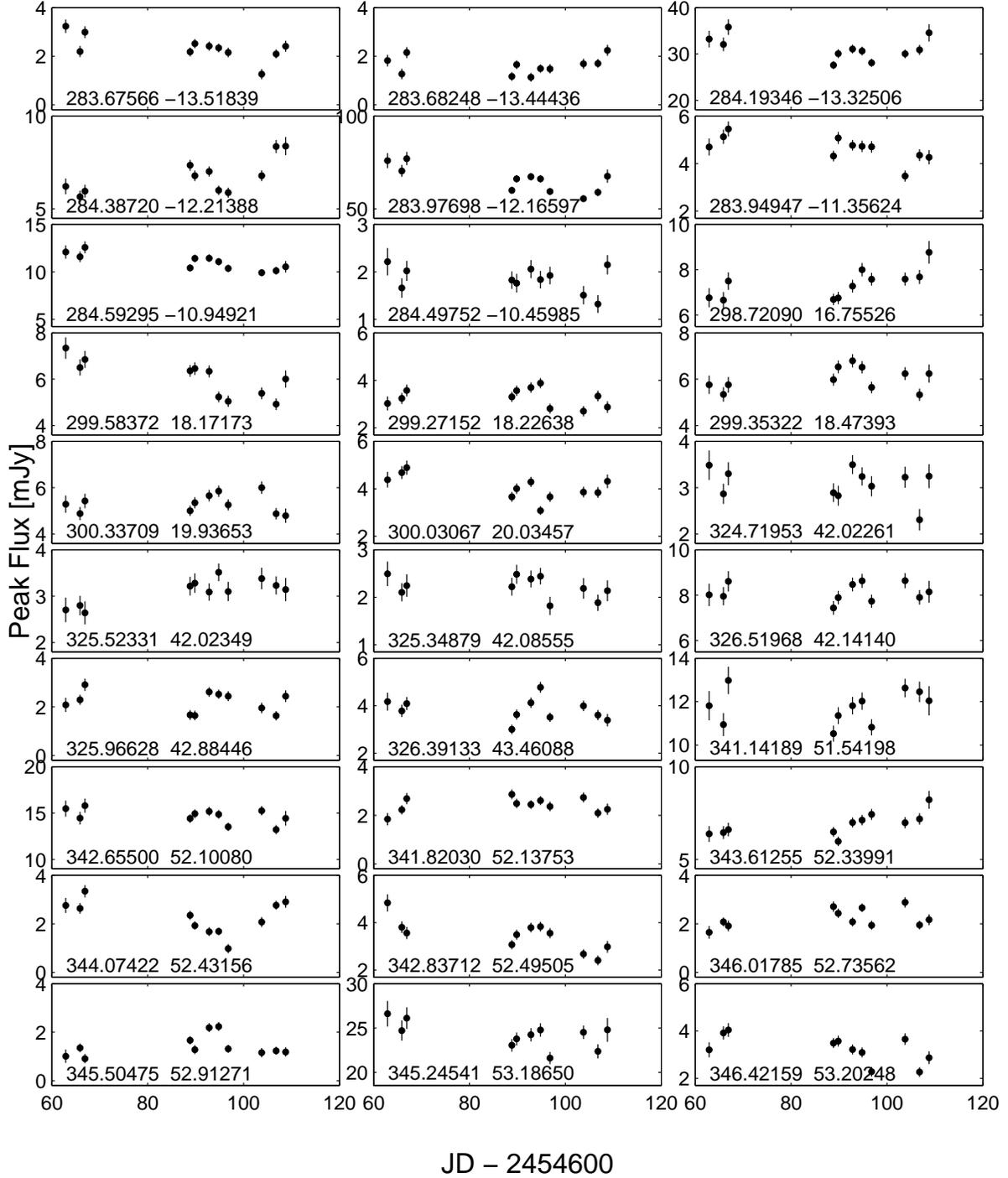}}
\caption{Light curves of the 30 variables which have $\chi^{2}>45.5$.
For scaling purposes we show only the 11 observations taken during 2008.
In each panel we give the J2000.0 Right Ascension and Declination of the source.
The individual flux measurements are given in the electronic version
of Table~\ref{Tab:MasterCat}.
\label{Fig:Var_LC}}
\end{figure*}

In total, we find that $30\%$ (30 out of 98) of the sources in our survey, which are
brighter than 1.5\,mJy, are variable
(at the 4-$\sigma$ level\footnote{Assuming Gaussian noise, $4\sigma$ corresponds to a probability of $\cong 1/15,000$ while the number of measurements in our experiment  (number of epochs multiplied by the number of sources) is $\approx 7400$.}).
The light curves of these 30 variable sources are presented in
Figure~\ref{Fig:Var_LC}, and their flux measurements and basic properties
are listed in Table~\ref{Tab:MasterCat}.
This is considerably larger than the
fraction of variables reported in some of the other ``blind'' surveys
listed in Table~\ref{Tab:PreviousSurveys}
(e.g. Gregory \& Taylor 1986; de Vries et al. 2004; Becker et al. 2010). 
A possible explanation for this apparent discrepancy is that the
sources in these surveys were extracted from mosaic images in which
each point in the survey footprint was observed multiple times
during several days.  Therefore, the fluxes they reported are average
fluxes over several days time scales (column $\delta{t}$ in Table~\ref{Tab:PreviousSurveys}).
Such measurements will tend to
average out variability on time scales which are shorter than $\delta{t}$.
In order to test this hypothesis we carried out a structure function analysis.

\begin{figure}
\centerline{\includegraphics[width=8.5cm]{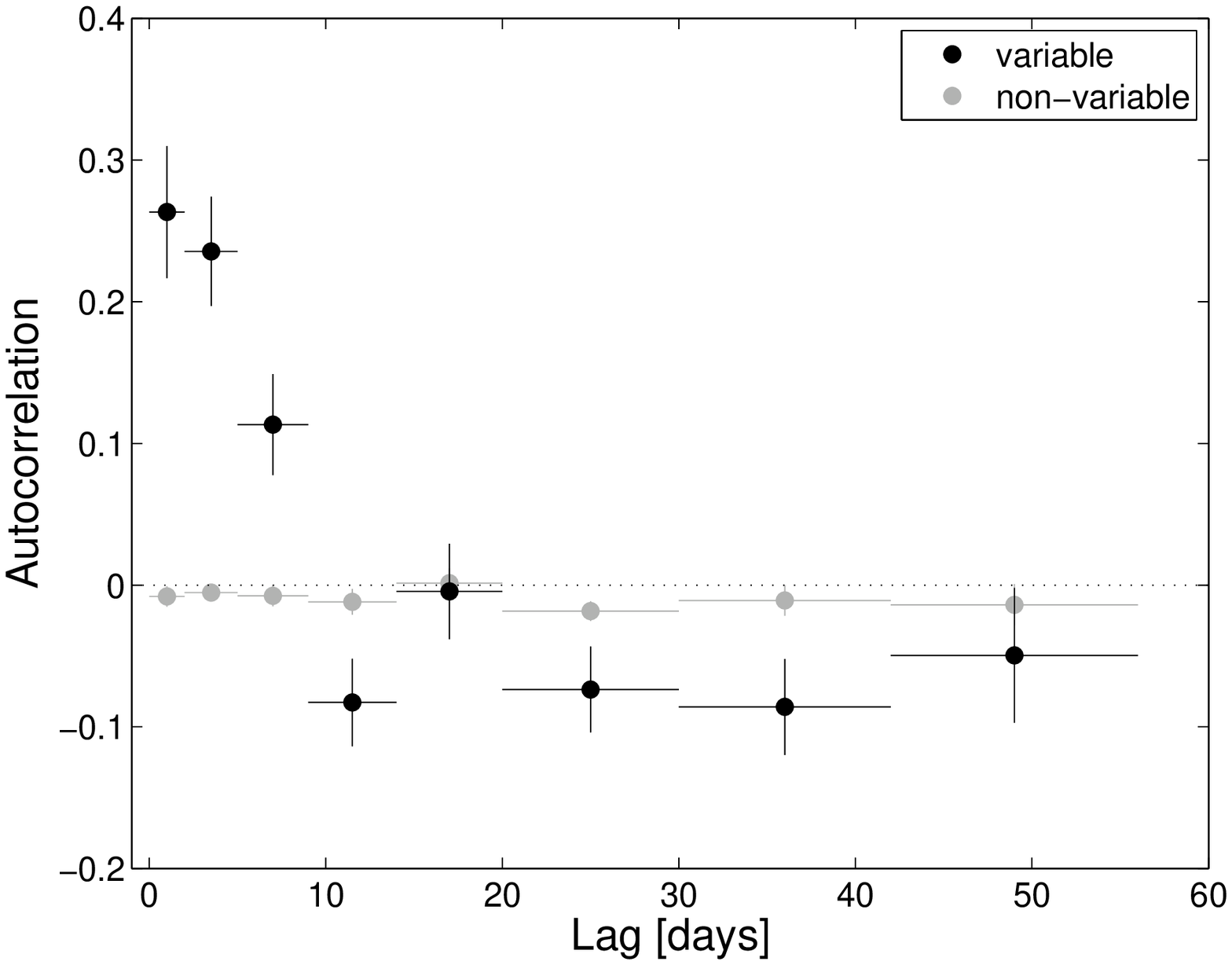}}
\caption{The mean discrete auto-correlation function of all the 30 variable sources (black circles).
The gray circles show the same but for all the non-variable sources
(see text).
Each light curve was normalized by subtracting its mean and dividing it by its mean.
The vertical error bars represent the $1$-$\sigma$ errors calculated
using the bootstrap method. The horizontal ``error bars'' represent the full width
of the lags contained within each bin.
The horizontal dotted line marks zero correlation.
\label{Fig:ACF_VarLC}}
\end{figure}

We calculated the mean discrete auto-correlation function, $C(\tau)$,
of all the 30 variable sources, as a function of the time lag $\tau$.
We first normalized each source light curve by subtracting its mean and than
dividing it by its (original) mean.
We treated all these light curves as a single light curve
by concatenating them with gaps, which are larger than the time span
of each light curve, in between light curves.
Then we followed the prescription of Edelson \& Krolik (1988) for calculating
the discrete auto-correlation function.
The errors were calculated using a bootstrap technique with 100 realizations
for the measurements in each time lag
(e.g., Efron 1982; Efron \& Tibshirani 1993).
The mean auto-correlation function is presented in Figure~\ref{Fig:ACF_VarLC} (black circles).
The auto-correlation at lag ``zero'' is not one. This is because at lag zero
we used a lag window of 0 to $+2$\,days.
Therefore it does~not contain only zero lag data.
The auto-correlation function reaches zero correlation at $\tau\approx10$\,days.

\begin{figure}
\centerline{\includegraphics[width=8.5cm]{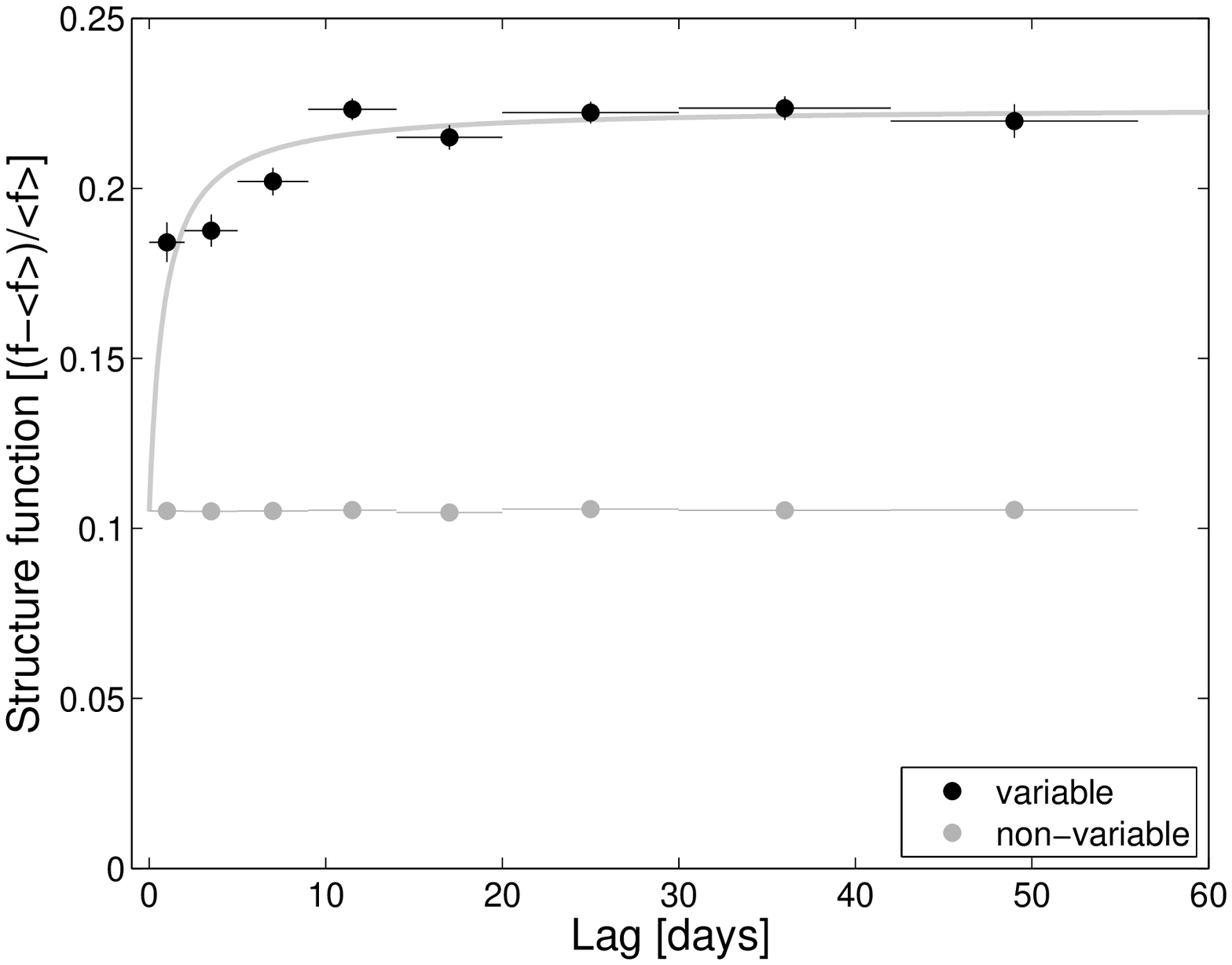}}
\caption{The mean structure function of the normalized light curves
as a function of lag, $\tau$.
Symbols like in Figure~\ref{Fig:ACF_VarLC}.
Following Lovell et al. (2008) we fit the structure function
(after subtracting the non-variable sources structure function)
with the function $m_{{\rm s}}\tau/(\tau+\tau_{{\rm char}})$.
The heavy gray line represent the best fit function,
and the best fit parameters are $m_{{\rm s}}=0.119\pm0.003$,
$\tau_{{\rm char}}=0.83_{-0.23}^{+0.27}$\,day ($\chi^{2}/dof=26.6/6$).
\label{Fig:SF_VarLC}}
\end{figure}

Next, we calculated the structure function, $SF(\tau)$,
of these light curves defined by
\begin{equation}
SF(\tau) = \sqrt{2S^{2}(1-C[\tau])} \pm \frac{2S^{2}\Delta{C[\tau]}}{2\sqrt{2S^{2}(1-C[\tau])}},
\label{Eq:SF}
\end{equation}
where $S$ is the standard deviation of the normalized light curves
and $\Delta{C[\tau]}$ is the bootstrap error in the discrete auto-correlation function.
The second term on the right hand side of Eq.~\ref{Eq:SF} represents
the error in the structure function.
The structure function is presented in Figure~\ref{Fig:SF_VarLC}.
Also shown in Figures~\ref{Fig:ACF_VarLC} and \ref{Fig:SF_VarLC} are
the auto-correlation and structure functions,
respectively, for
``non-variable'' sources (gray symbols).
In this context our selection criteria for non-variable sources
are specific flux larger than 2\,mJy and $\chi^{2}<25.3$.
This $\chi^{2}$ value corresponds to 2-$\sigma$ confidence, assuming 15 degrees of freedom.

The structure function of the variable sources, after subtracting the
non-variable source
structure function is rising rapidly 
from zero to $\approx0.05$ on a time scale of the
order of one day and than rises to a level
of $\approx0.12$ at lags of $\approx10$\,days
at which it stays roughly constant.
Though, we cannot rule out it is slowly rising on $\tau>10$\,days time scales.

This analysis suggests that a large component of the variability
happens on time scales shorter than about one day.
A plausible explanation is that the short time scale variability
is due to refractive scintillations and it is discussed in \S\ref{VarDisc}
(e.g., Rickett 1990).
This level of
variability was likely missed by some previous surveys (Table~\ref{Tab:PreviousSurveys})
due either to their choice of frequency ($\nu$), cadence (N$_{ep}$) or the
observing time span ($\delta{t}$). On the other hand, 5\,GHz surveys of
flat spectrum active galactic nuclei (AGN) find that their majority
shows significant variability, and that the
number of these variable sources increase at low Galactic latitudes
(e.g., Spangler et al. 1989;
Ghosh \& Rao 1992;
Gaensler \& Huntstead 2000;
Ofek \& Frail 2011).
The source population in the flux density range of our
survey is known to be dominated by AGN and a significant fraction
($\sim50$\%) of these are compact, flat-spectrum AGN and hence
expected to show short-term flux density variations (de Zotti et al. 2010).

\subsection{Variability on years time scale}
\label{VarYear}

By comparing our 2008 and 2010 catalogs we were able to
probe the variability of sources brighter than $\gtorder 0.5$\,mJy
on two year time scale.
However, this comparison is limited to only two epochs.
Each of the two epochs in our two year time-scale variability analysis
is composed of multiple epochs taken ($\approx\delta{t}$) days to weeks apart.
Therefore, in this analysis short-term variability
is averaged out.
For example, in 5\,GHz, variability due to scintillations will
typically have time scale which is shorter than a few days.

\begin{figure*}
\centerline{\includegraphics[width=16cm]{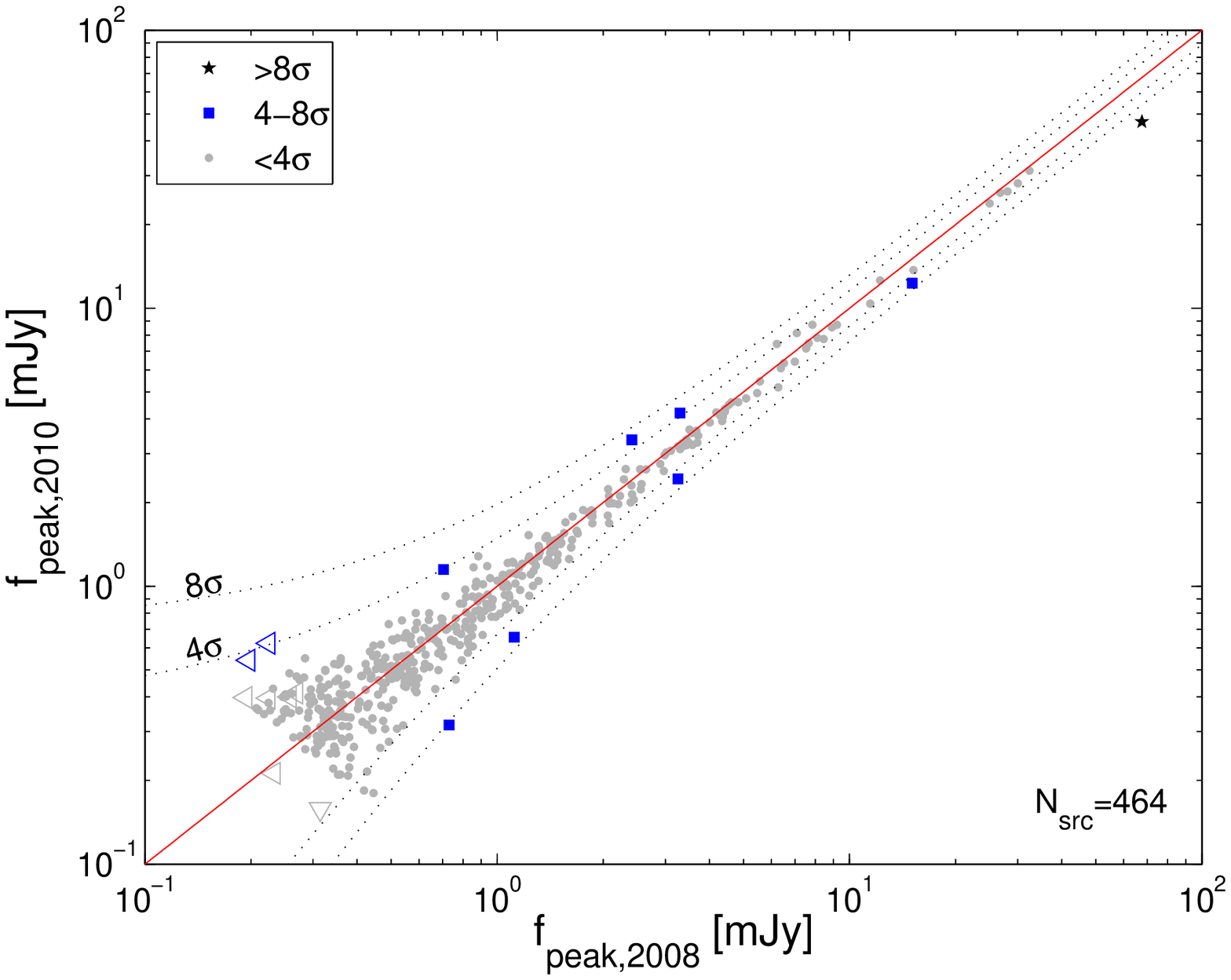}}
\caption{A comparison of the 2008 and 2010 peak flux densities for all
  radio sources in the Master catalog. Equal fluxes are indicated by a
  solid line.
  The dashed lines are the
  4$\sigma$ and 8$\sigma$ confidence levels contours
  based on the mean noise properties
  of the images and are calculated assuming the noise has a constant
  component, which is the rms of the two yearly
  images (Table~\ref{Tab:SurveyPar})
  added in quadrature with a flux dependent component
  which is given by the cosmic errors discussed in \S\ref{PostSurveyCat}.
  The average rms noise
  for the summed 2008 images was 72\,$\mu$Jy\,beam$^{-1}$ and for 2010
  images it was 56\,$\mu$Jy\,beam$^{-1}$.
  The shape and color coding represent the $\chi^{2}$ (one $dof$)
  of individual objects
  based on the actual noise rather than the average noise (see legend).
  Open triangles mark limits were the direction of the limit
  is given by the direction of the tip of the triangle (bottom or left).
\label{Fig:Flux2008vsFlux2010}}
\end{figure*}

Figure~\ref{Fig:Flux2008vsFlux2010} shows
a comparison of the 2008 and 2010 peak flux densities for all
radio sources in the Master catalog.
The dashed lines represent the mean noise 4 and 8-$\sigma$ confidence
variability contours.
Based on the $\chi^{2}_{Y}$ (Table~\ref{Tab:MasterCat})
we find ten
sources (2\% of all the sources) which have variability with confidence level larger than
about 4$\sigma$ (i.e., $\chi^{2}_{Y}>16$; assuming one degree of freedom).
These variable candidates are listed in Table~\ref{Tab:MasterCat} (below the horizontal line).
We note that the $\chi^{2}_{Y}$ calculation takes into account
the gain correction factors and the cosmic errors described in \S\ref{PostSurveyCat}.
We visually inspected the images of all ten variable candidates
and verified they are not sidelobe artifacts.

J$213626.36+413926.6$ is the only significant strong variable (i.e.
with $\chi^{2}/dof>16/1$ and $V_{{\rm F}}>0.5$) in our two-year comparison.
In 2010 it has a peak flux density (uncorrected for primary beam attenuation) of
$627\pm50$\,$\mu$Jy ($12.5\sigma$),
while in 2008 there is a nominal detection of
the source at this position with a peak flux density of $197\pm75$\,$\mu$Jy ($2.6\sigma$).
The source flux was well below the detection threshold
for every one of the 11 epochs in 2008 but it is visible in all five
epochs in 2010.
We note that this source is one of the two faint sources seen above
the dashed line on the right side of Figure~\ref{Fig:VarEp16_StDvsChi2}.
For this reason we do not classify this as a transient source
but it appears instead to be a persistent radio source that has
tripled its flux density over a two year interval. There is no known
cataloged source at this position in the NVSS catalog (Condon et al. 1998)
nor in the SIMBAD or NED databases.

Our Master catalog contains 317 sources brighter than J$213626.36+413926.6$.
Therefore, we roughly estimate that the fraction of strong variables
with two epochs $V_{{\rm F}}>0.5$ is $0.32_{-0.26}^{+0.73}\%$.
However, we note that this measurement is based on averaging out
variability on time scales shorter than a few weeks.

\section{Discussion}
\label{Disc}

We present a 16 epoch, Galactic plane survey, for 
radio transients and variables at 5\,GHz.
We detected one possible transient and many variable
sources.
The transient areal density and rate based on this
single detection are derived in \S\ref{TranRate}.
In \S\ref{VarDisc} we discuss our variability
study and compare it with previous surveys.

\subsection{Transient areal density and rate}
\label{TranRate}

The analysis of our data revealed a single radio transient candidate.
The area encompassed within a single half power beam (radius of $r_{{\rm HP}}=4.65'$)
in which we searched for transients
is $0.00601$\,deg$^{2}$ and the area we targeted within the half power radius of
all 141 fields is $2.66$\,deg$^{2}$.
Given that each field was observed on average 15.70 times
(sum of column four in Table~\ref{Tab:ListOfFields}
divided by the number of fields),
the total area covered by our survey over all the epochs is $41.2$\,deg$^{2}$.
Our survey used an uncorrected flux limit of 1.5\,mJy for transients.
However, the sensitivity within the field of view of a single beam
imaging is not uniform, and degrades by a factor of two at the half power radius.
In order to calculate the transient areal density
from these parameters
we need to assume something about the source number count function.
We parametrize the source number count function as a power law
of the form
\begin{equation}
\kappa(>f)=\kappa_{0}(f/f_{0})^{-\alpha},
\label{Eq:CountFun}
\end{equation}
where $f$ is specific flux, $\kappa(>f)$ is the sky surface density of sources
brighter than $f$,
$\kappa_{0}$ is the sky surface density of sources brighter than $f_{0}$,
and $\alpha$ is the power law index of the source number count function.
It is well known that for homogeneous source distribution in
an Euclidean universe
and arbitrary luminosity function $\alpha=3/2$.
In Appendix~\ref{Ap:Rate} we derive a simple relation
for the number of sources that are expected
to be detected in a beam with
a power sensitivity that falls like a Gaussian as a function
of $f_{0}$, $\kappa_{0}$, $\alpha$, $r_{{\rm HP}}$,
the search radius $r_{{\rm max}}$, and the specific flux
limit at the beam center $f_{{\rm min},0}$.

Based on Equation~\ref{Eq:kappa},
and assuming $\alpha=3/2$, we find that
the transient areal density at 1.8\,mJy
(corrected for the CLEAN bias) is
\begin{equation}
\kappa(>1.8\,{\rm mJy}) = 0.039_{-0.032,-0.038}^{+0.13,+0.18}\,{\rm deg}^{-2},
\label{Eq:SurveyArealDen}
\end{equation}
where the errors correspond to one and two sigma confidence intervals,
calculated using the prescription of Gehrels (1986).
If our detected transient is not real then our
survey poses a 95\% confidence upper limit
on the transient rate of $\kappa(>1.8\,{\rm mJy})<0.15$\,deg$^{-2}$.

Translation of our areal density to transient rate
depends on the transient duration $t_{{\rm dur}}$ and it is
\begin{equation}
\Re(>1.8\,{\rm mJy}) = (28_{-23,-27}^{+65,+132}) \Big( \frac{t_{{\rm dur}}}{0.5\,{\rm day}} \Big)^{-1}\,{\rm deg}^{-2}\,{\rm yr}^{-1}.
\label{Eq:SurveyTranRate}
\end{equation}
Note that this translation is correct only if $t_{{\rm dur}}$ is smaller than
the time between epochs.

Figure~\ref{Fig:ArealDen_Flux_SurveySummary}
presents a summary of the radio transient and persistent source areal density
as observed by various searches and at different frequencies.
This figure is largely based on Table~\ref{Tab:PreviousSurveys}
and the areal density reported here.
As shown in this figure, the areal density derived in this
work is consistent with the expectation based on
the Bower et al. (2007) transient sky surface density.
Moreover, it is roughly consistent with the sky
surface density of the Nasu survey transients.
We note that the Nasu sky surface density
is based on our limited knowledge of this project
(see \S\ref{Prev}).
This comparison assumes that the transient areal density
on the celestial sphere is uniform.

This figure implies that the areal density of radio
transients in the sky is roughly 2--3 orders of magnitude below
the persistent radio source sky surface density.
This is in contrast to the visible light sky
in which the fraction of transients (excluding solar system minor planets)
among persistent sources is roughly $10^{-4}$ down
to the limiting magnitude of surveys like
the Palomar Transient Factory (Law et al. 2010; Rau et al. 2010).

It is interesting to compare this figure with some recent predictions.
Nakar \& Piran (2011) predict that compact binary mergers,
regardless whether they are associated with short-duration gamma-ray bursts
(e.g., Nakar 2007),
will produce radio afterglows with a duration of several months.
They suggest that the two-months long
radio transient RT19870422 detected by Bower et al. (2007)
may be a binary merger radio afterglow.
Moreover, they find that the rate inferred from this
event is consistent with the predicted rate
of binary merger events.

Giannios \& Metzger (2011) suggested that tidal flare events
may produce radio transients with durations of month to years,
with a 5\,GHz peak flux of 1\,mJy at a distance of 1\,Gpc (i.e., $z\cong0.2$).
The total comoving volume\footnote{Assuming WMAP fifth year cosmological
parameters (Komatsu et al. 2009).}
enclosed within a luminosity distance of 1\,Gpc is $2.4\times10^{9}$\,Mpc$^{3}$
or $5.9\times10^{4}$\,Mpc$^{3}$\,deg$^{-2}$.
Bower et al. (2007) did~not find any
transients with a duration of two months
which are associated with the nucleus of a galaxy.
This is translated to a 95\% confidence upper limit on
the rate of radio tidal flare events, brighter than 1\,mJy,
of $\sim0.1$\,deg$^{-2}$\,yr$^{-1}$.
Therefore, in the context of the Giannios \& Metzger (2011)
predictions,
we can put an upper limit of $\sim7\times10^{-6}$\,Mpc$^{3}$\,yr$^{-1}$
on the rate of tidal flare event radio afterglows.
This is in rough agreement with
the predicted tidal flares rate (Magorrian \& Tremaine 1999; Wang \& Merritt 2004).

Finally, we note that our detection rate is consistent with
an old NS origin as suggested in Ofek et al. (2010),
and with their expected surface density at low Galactic latitude
based on the Ofek (2009) simulations.

\subsection{Comparison of variability with previous surveys}
\label{VarDisc}

We analyzed the variability of the sources detected in our survey on
days--weeks time scales and two years period.  On short time
scales we find that considerable fraction of the bright point sources
are variable.  At the 10--100\,mJy range it seems that more than half
the sources are variable on some level ($>4\sigma$).
Furthermore, we find that
30\% of the sources in our survey, which are brighter than
$\approx1.5$\,mJy,
are variable (at the 4-$\sigma$ level).  This is considerably higher
than the fraction of variables reported in some other surveys (e.g. Gregory
\& Taylor 1986; de Vries et al. 2004; Becker et al. 2010).
We suggest that a possible explanation for this apparent discrepancy is
that the sources in these surveys were extracted in a way that
washed out short time scale variability (see \S\ref{VarSingle}).
This is supported by the fact that our two years time scale
variability study, in which each epoch is composed by averaging
multiple observations, shows smaller fraction of variables.
Moreover, our structure function analysis shows that a big fraction of
the variability component happens on time scales shorter than about 
a few days.
We note that these fast variation of radio sources
is known for a long time, and was found
by Heeschen (1982; 1984).
Moreover, a large fraction of variable sources was previously reported by
some other efforts, which did~not average out short time
scales variability (e.g., Lovell et al. 2008).

We speculate that the fast rise of the structure function on 
$\approx10$\,day 
time scales is due to scintillations in the interstellar
medium (ISM).
These time scales are consistent
with those expected theoretically from refractive scintillations in 5\,GHz
(e.g., Blandford et al. 1986; Hjellming \& Narayan 1986).
Moreover, similar rise times were reported by other efforts
(e.g., Qian et al. 1995).
Unlike diffractive scintillation which may
produce strong variability
(StD/$\langle f\rangle$ of $80\%$; e.g., Goodman 1997; Frail, Waxman \& Kulkarni 2000),
refractive scintillations can easily explain the observed
amplitude of $\approx13\%$.
We note that a comparison of models with observations
suggests that most of the radio source variability below
5\,GHz is due to scintillations, while above 5\,GHz
there is an intrinsic variability component
(e.g., Hughes, Aller \& Aller 1992; Mitchell et al. 1994; Qian et al. 1995)
Therefore, we can not rule out the possibility that some of
the variability we detected in our survey is intrinsic to the sources.

After averaging out variations on days time scale, our two year
variability analysis indicates that about $0.3\%$ of the sources above
0.5\,mJy are strong variables ($V_{{\rm F}}>0.5$ for $N_{{\rm ep}}=2$),
and that only a small
fraction $\approx3\%$ of the sources are variables at some level.  This
finding supports the hypothesis that the main reason for low-amplitude
radio variability at 5\,GHz is due to scintillations.

It is well known that the fraction of radio variable sources
increases toward the Galactic plane
(e.g., Spangler et al. 1989; Ghosh \& Rao 1992;
Gaensler \& Hunstead 2000; Lovell et al. 2008; Becker et al. 2010; Ofek \& Frail 2011),
plausibly due to Galactic scintillations.
However, there are some
claims as yet unconfirmed that the number of 
{\it intrinsically} strong variables, at 5\,GHz, increases toward the Galactic plane
(Becker et al. 2010).
Specifically, Becker et al. (2010) suggested that there is a
separate Galactic population
of {\it strong} variables. 
As noted in \S\ref{Prev}, they found more
than half of their variables, on timescales of
years, varied by more than 50\% in the 1-100 mJy flux density range.
Moreover, they found that these strong variables were concentrated at low Galactic
latitudes and toward the inner Galaxy.
In contrast, we find a much smaller fraction
of strong variables (see \S\ref{VarYear}).
However, Becker et
al. (2010) observed sources within one degree of the Galactic plane,
while our survey sampled Galactic latitudes $\vert b\vert \cong6$--$8^\circ$.

\section{Summary}
\label{Sum}

We present a VLA 5\,GHz survey to search for radio transients
and explore radio variability in the Galactic plane.
Our survey represent the first attempt to discover radio transients in 
near real time and initiate multi-wavelength followup.
Our real time search identified two possible transients.
However, followup observations and our post-survey analysis
showed that these candidates are not transient sources.
Nevertheless, in one case, we were able to initiate visible light observations
of the transient candidate field only one hour after the candidate
was detected.

Our post survey analysis reveals one possible transient source
detected at the 5.8-$\sigma$ level.
Our P60 images of this transient field, taken two days after
the transient detection, do~not reveal any visible light source
brighter than $i$-band magnitude of 21
associated with the transient within $2''$.
The transient has a time scale longer than 1\,min.
However, we cannot put an upper limit on its duration
since it was detected on the first epoch of our survey.
Based on this single detection we find a transients brighter than 1.8\,mJy areal density
of $\approx0.04$\,deg$^{-2}$.
The transient surface density found in this paper is compared with other surveys
in Figure~\ref{Fig:ArealDen_Flux_SurveySummary}.
Our transient areal density is consistent with the one reported
by Bower et al. (2007), corrected for the flux limit.
This is also roughly consistent, up to a possible spectral correction
factor, with the rates reported by Levinson et al. (2002) and Kida et al. (2007).
Finally, based on existing evidence, we cannot rule out
the hypothesis that these transients are
originating from Galactic isolated old NSs (Ofek et al. 2010).

Finally, we present a comprehensive variability analysis of our data,
with emphasis on proper calibration of the data and estimating
systematic noise.
Our findings suggest that short time scale variability
among 5\,GHz point sources is common.
In fact above 1.5\,mJy at least 30\% of the point sources
are variable with variability exceeding our 4-$\sigma$ detection level.
This is consistent with the Lovell et al. (2008) results,
and is plausibly explained by refractive scintillations in the ISM.

\acknowledgments

We are grateful to Barry Clark for his
help with scheduling the VLA observations.
EOO is supported by an Einstein fellowship and NASA grants.
Support for program number HST-GO-11104.01-A was provided by NASA through
a grant from the Space Telescope Science Institute, which is
operated by the Association of Universities for Research in
Astronomy, Incorporated, under NASA contract NAS5-26555.
AG acknowledges support
by the Israeli Science Foundation, an EU Seventh
Framework Programme Marie Curie IRG fellowship, the
Benoziyo Center for Astrophysics, and the Yeda-Sela fund
at the Weizmann Institute.

\appendix
\section{Estimate of the gain correction factors}
\label{Ap:ZP}

In order to construct the best light curves of the sources
we need to remove any systematic factors influencing
the measurements. A way to do this is to use the fact that
the actual light curves of many sources are not correlated.
Therefore, in the absence of systematic factors affecting
the measurements, the scatter in the average light curve
should be minimized.
In order to minimize the scatter in the average light curve of our sources
we multiplied the fluxes at each epoch of observation (taken during
$\sim3$~hours) by a ``gain'' correction factor,
such that the residuals in all light curves, compared
with a constant light curves, will be minimized.  This problem, is
similar to producing relative photometry light curves in optical
astronomy.

We used a linear least square minimization technique.  Using
this method we are solving for the best zero point normalization (per
epoch), and the best ``mean'' flux of each source, that minimize the
global scatter in all the light curves.
This technique was already introduced
by Honeycutt (1992), but here we write it in
a more easy to use form and we also
add linear set of constraints for simultaneous absolute calibration.

We work in a ``magnitude'' system
(i.e., $m = -2.5\log_{10}{f}$).
The advantage of the magnitude system is that the
logarithm of a quantity has the property of making error distribution
more symmetrical, and this may be somewhat important
for faint sources.
The basic idea of this technique is to simultaneously
solve the following set of equations in the least square sense:
\begin{equation}
m_{ij} \cong z_{i} + \bar{m}_{j},
\label{Eq:BasicZP}
\end{equation}
where $m_{ij}$ is a $p\times q$ matrix that contains
all the measured (``instrumental'') magnitudes,
$i$ is the epoch index ($p$ epochs), and $j$ is the source index
($q$ sources).
Here, 
$\bar{m}_{j}$ is the mean magnitude of the $j$-th source
and $z_{i}$ is the zero point of the $i$-th epoch.
We note that $\bar{m}_{j}$ and $z_{i}$ are free parameters.
In case we have error measurements,
let $\sigma^{m}_{ij}$ be the respective errors in
the instrumental magnitudes.
In some cases, we may have
additional constraints like the calibrated magnitudes, $M_{j}$,
(and respective errors, $\sigma^{M}_{j}$)
of some or all the sources.
This additional information can be used as constraints
on the system of linear equations.
Using these constraints our output magnitudes will be
calibrated in respect to a set of reference sources.

Given $m_{ij}$ and $\sigma^{m}_{ij}$,
and the optional $M_{j}$ and $\sigma^{M}_{j}$
we would like to find the properly weighted best fit free parameters
$z_{i}$ and $\bar{m}_{j}$.
Let $\vec{m}$ be a vector of the observable quantities
(reorganization of $m_{ij}$),
and $\vec{\sigma}$ the respective vector of errors in these quantities
obtained by re-arranging the matrices of instrumental 
and calibrated magnitudes (i.e., $m_{ij}$ and $\sigma^{m}_{ij}$):
\begin{equation}
\begin{array}{ll}

\vec{m} = \left[ \begin{array}{c}
         m_{11} \\
         m_{12} \\
         \vdots  \\
         m_{1q} \\ \hline
         m_{21} \\
         m_{22} \\
         \vdots  \\
         m_{2q} \\ \hline
         \vdots  \\ \hline
         m_{p1} \\
         m_{p2} \\
         \vdots   \\
         m_{pq} \\ \hline \hline
         M_{1}   \\
         M_{2}   \\
         \vdots  \\
         M_{q}   \\     
    \end{array} \right],  &

\vec{\sigma} = \left[ \begin{array}{c}
         \sigma^{m}_{11} \\
         \sigma^{m}_{12} \\
         \vdots  \\
         \sigma^{m}_{1q} \\ \hline
         \sigma^{m}_{21} \\
         \sigma^{m}_{22} \\
         \vdots  \\
         \sigma^{m}_{2q} \\ \hline
         \vdots  \\ \hline
         \sigma^{m}_{p1} \\
         \sigma^{m}_{p2} \\
         \vdots   \\
         \sigma^{m}_{pq} \\ \hline \hline
         \sigma^{M}_{1}   \\
         \sigma^{M}_{2}   \\
         \vdots  \\
         \sigma^{M}_{q}   \\     
    \end{array} \right].
\end{array}
\label{Eq:Y}
\end{equation}
Note that the elements below the double horizontal line
are optional elements, needed only if we want to simultaneously apply
a magnitude calibration.

Next, we can define a vector of free parameters
we would like to fit:
\begin{equation}
\vec{P} = \left[ \begin{array}{ccccccccc}
         z_{1} & z_{2} & ... & z_{p} & , & \bar{m}_{1} & \bar{m}_{2} & ... & \bar{m}_{q} \\
    \end{array} \right]^{T},
\label{Eq:P}
\end{equation}
where the superscript $T$ indicate a transpose operator.
In that case the design matrix $H$, which satisfy
$\vec{m}\cong H\vec{P}$
is easy to construct:
\begin{equation}
H = \left[ \begin{array}{c|c}
       I_{j=1}^{q\times p} & I^{q\times q} \\  \hline
       I_{j=2}^{q\times p} & I^{q\times q} \\  \hline
       \vdots              & I^{q\times q} \\  \hline
       I_{j=q}^{q\times p} & I^{q\times q} \\  \hline\hline
       0^{q\times p}       & I^{q\times q} \\
    \end{array} \right],
\label{Eq:H}
\end{equation}
where $I_{j=k}^{q\times p}$, is a $q \times p$ matrix
in which the $k$-th column contains ones, while
the rest of the elements are zeros,
$I^{q\times q}$ is a $q\times q$ identity matrix,
and $0^{q\times p}$ is a $q\times p$ zeros matrix.

Note that (again) the lower block of the matrix $H$, separated
by two horizontal lines, is an optional section
that is used for the magnitude zero-point calibration.
This additional section acts like constraints
on the system of linear equations.

The rank of the matrix $H$ without the lower block is $p+q-1$.
This is because without the calibration block there is
an arbitrariness in adding a zero-point to each epoch.
Adding the calibration block (or part of it)
fixes this problem and in that
case the rank of $H$ is $p+q$.
In cases when a given source does~not appear in a specific
epoch, we simply have to remove the appropriate
row in $H$, $\vec{m}$ and $\vec{\sigma}$.

In order to find the best fit parameters $\vec{P}$,
and their respective errors $\sigma^{P}$,
we need to find $\vec{P}$ that minimizes the $\chi^{2}$:
\begin{equation}
\chi^{2} = (\vec{m} - H\vec{P})^{T} [\sigma_{ij}^{2}]^{-1} (\vec{m}-H\vec{P}),
\label{Eq:Chi2}
\end{equation}
where $[\sigma^{2}_{ij}]$ is the matrix of measurement errors
$\sigma^{m}_{ij}$.
The problem of finding $\vec{P}$ and their corresponding errors is described
in many textbooks
(e.g., Press et al. 1992; for a tutorial see Gould 2003).

The design matrix, $H$, even without the magnitude calibration part,
is an $(p\times q)\times (p + q)$ matrix.
For many problems, the matrix $H$, may be huge and requires
a lot of computer memory.
However, $H$ is highly sparse, and therefore sparse matrix
utilities can be used if needed.
Alternatively, this can be solved 
using the conjugate gradient
method\footnote{See basic description and overview in: http://www.cs.cmu.edu/$\sim$quake-papers/painless-conjugate-gradient.pdf}.

We note that in practice this method may be applied iteratively.
After the first iteration, we can check
the $\chi^{2}_{j}$ for each source and 
the $\chi^{2}_{i}$ for each epoch.
Then, we can remove sources with large value of $\chi^{2}_{j}$,
and/or we can add a ``cosmic'' errors to all the measurements
in an epoch with large $\chi^{2}_{i}$.
After we construct the new $H$ and $Y$ we may
apply the inversion again.

We note that typically, in
addition to $\sigma_{ij}^{m}$ (the errors associated with
the individual sources)
there are additional errors
(e.g., calibration errors in radio astronomy and flat fielding
errors in optical astronomy).
Ignoring these errors is not recommended since it
will give over weight to sources with small
errors. A solution to this problem is, again, 
to apply this method in iterations.
After the first iteration it is possible to estimate
the cosmic error term (based on the residuals from the best fit),
and to add these cosmic errors to the instrumental errors in the second
iteration.

Finally, using this method additional de-trending is possible.
For example, one can add additional columns to the design matrix
(and corresponding additional terms to the vector of free parameters)
that represent changes in the zero point as a function
of additional parameters.
For example, in radio astronomy, this may be a variation
in the zero point as a function of the distance from the beam
center, and in optical astronomy this may be an airmass-color term,
positional terms, color terms affecting different instruments and more.

\section{$V_{{\rm R}}$ and $V_{{\rm F}}$ statistics}
\label{Ap:StatVR}

The $V_{{\rm R}}$ and $V_{{\rm F}}$ defined in Equations~\ref{VR}--\ref{VF}
are sensitive to the number of measurements.
This complicates the comparison between surveys with different number of epochs.
In order to demonstrate this and to provide with
a mean for roughly convert these variability indicators between different surveys,
here we calculate the expectation value for $V_{{\rm R}}$ as a function
of the number of epochs in a survey, $N_{{\rm ep}}$, and the StD/$\langle f\rangle$
of a source light curve.
We note that $V_{{\rm R}}$ and $V_{{\rm F}}$ are exchangeable.

In order to calculate this conversion we performed the following simulations.
We generated random light curves with $N_{{\rm ep}}$ points
and which are drawn from a log-normal standard deviation, $S_{{\rm logN}}$.
For each value of $N_{{\rm ep}}$ and $S_{{\rm logN}}$
$10^{6}$ light curves were generated and
StD/$\langle f\rangle$ and $\langle V_{{\rm R}}\rangle$
were calculated.
In Table~\ref{Tab:StatVR} we list
the $V_{{\rm R}}$ expectation values as a function of $N_{{\rm ep}}$ (rows)
and StD/$\langle f\rangle$ (columns).
Below the StD/$\langle f \rangle$ we give also the appropriate 
$S_{{\rm logN}}$.
\begin{deluxetable*}{lllllllllllllllllll}
\tablecolumns{19}
\tabletypesize{\scriptsize}
\tablewidth{0pt}
\tablecaption{Translation of $\langle V_{{\rm R}}\rangle$ to StD/$\langle f\rangle$}
\startdata
\hline
\hline
StD/$\langle f\rangle$ &  $0.01$& $0.05$& $0.10$& $0.15$& $0.20$& $0.25$& $0.31$& $0.35$& $0.38$& $0.42$& $0.43$& $0.44$& $0.45$& $0.46$& $0.47$& $0.48$& $0.50$& $0.51$ \\
\vspace{1.5mm} 
$S_{{\rm logN}}$         &  $0.01$& $0.05$& $0.10$& $0.15$& $0.20$& $0.25$& $0.30$& $0.34$& $0.37$& $0.40$& $0.41$& $0.42$& $0.43$& $0.44$& $0.45$& $0.46$& $0.47$& $0.48$ \\

\hline

$N_{{\rm ep}}$&     &       &       &       &       &       &       &       &        &      &       &       &       &       &       &       &       &        \\
\vspace{1.5mm} 
         2&  $1.01$& $1.06$& $1.12$& $1.19$& $1.27$& $1.36$& $1.45$& $1.54$& $1.60$& $1.68$& $1.70$& $1.73$& $1.75$& $1.78$& $1.81$& $1.83$& $1.86$& $1.89$ \\
\vspace{1.5mm} 
         3&  $1.02$& $1.09$& $1.19$& $1.30$& $1.43$& $1.57$& $1.72$& $1.87$& $1.98$& $2.11$& $2.15$& $2.20$& $2.24$& $2.29$& $2.34$& $2.39$& $2.44$& $2.49$ \\ 
\vspace{1.5mm} 
         4&  $1.02$& $1.11$& $1.23$& $1.37$& $1.53$& $1.72$& $1.92$& $2.11$& $2.27$& $2.43$& $2.49$& $2.55$& $2.62$& $2.68$& $2.75$& $2.81$& $2.88$& $2.96$ \\ 
\vspace{1.5mm} 
         5&  $1.02$& $1.12$& $1.27$& $1.43$& $1.62$& $1.83$& $2.08$& $2.31$& $2.50$& $2.70$& $2.77$& $2.85$& $2.92$& $3.00$& $3.09$& $3.17$& $3.26$& $3.35$ \\ 
\vspace{1.5mm}
         6&  $1.03$& $1.14$& $1.29$& $1.47$& $1.69$& $1.93$& $2.21$& $2.47$& $2.69$& $2.93$& $3.01$& $3.10$& $3.19$& $3.28$& $3.38$& $3.48$& $3.58$& $3.68$ \\ 
\vspace{1.5mm}
         7&  $1.03$& $1.15$& $1.32$& $1.51$& $1.74$& $2.01$& $2.32$& $2.61$& $2.86$& $3.13$& $3.22$& $3.32$& $3.42$& $3.53$& $3.64$& $3.75$& $3.87$& $3.98$ \\ 
\vspace{1.5mm}
         8&  $1.03$& $1.15$& $1.33$& $1.54$& $1.79$& $2.08$& $2.42$& $2.74$& $3.01$& $3.31$& $3.41$& $3.52$& $3.63$& $3.74$& $3.87$& $3.99$& $4.12$& $4.26$ \\ 
\vspace{1.5mm}
         9&  $1.03$& $1.16$& $1.35$& $1.57$& $1.84$& $2.15$& $2.51$& $2.86$& $3.14$& $3.46$& $3.58$& $3.70$& $3.82$& $3.95$& $4.08$& $4.22$& $4.36$& $4.50$ \\ 
\vspace{1.5mm}
        10&  $1.03$& $1.17$& $1.36$& $1.60$& $1.87$& $2.20$& $2.59$& $2.96$& $3.27$& $3.61$& $3.74$& $3.86$& $3.99$& $4.13$& $4.28$& $4.42$& $4.57$& $4.73$ \\ 
\vspace{1.5mm}
        11&  $1.03$& $1.17$& $1.38$& $1.62$& $1.91$& $2.26$& $2.66$& $3.05$& $3.38$& $3.75$& $3.88$& $4.01$& $4.15$& $4.30$& $4.45$& $4.61$& $4.77$& $4.94$ \\ 
\vspace{1.5mm}
        12&  $1.03$& $1.18$& $1.39$& $1.64$& $1.94$& $2.30$& $2.73$& $3.14$& $3.49$& $3.87$& $4.01$& $4.16$& $4.31$& $4.46$& $4.62$& $4.79$& $4.96$& $5.14$ \\ 
\vspace{1.5mm}
        13&  $1.03$& $1.18$& $1.40$& $1.66$& $1.97$& $2.35$& $2.80$& $3.22$& $3.58$& $3.99$& $4.13$& $4.29$& $4.45$& $4.61$& $4.78$& $4.96$& $5.14$& $5.33$ \\ 
\vspace{1.5mm}
        14&  $1.03$& $1.19$& $1.41$& $1.68$& $2.00$& $2.39$& $2.86$& $3.30$& $3.68$& $4.10$& $4.25$& $4.41$& $4.58$& $4.75$& $4.93$& $5.12$& $5.31$& $5.50$ \\ 
\vspace{1.5mm}
        15&  $1.04$& $1.19$& $1.42$& $1.69$& $2.03$& $2.43$& $2.91$& $3.37$& $3.76$& $4.21$& $4.36$& $4.53$& $4.70$& $4.88$& $5.07$& $5.27$& $5.47$& $5.68$ \\ 
\vspace{1.5mm}
        16&  $1.04$& $1.19$& $1.43$& $1.71$& $2.05$& $2.46$& $2.96$& $3.44$& $3.84$& $4.30$& $4.47$& $4.64$& $4.82$& $5.01$& $5.20$& $5.41$& $5.62$& $5.83$ \\ 
\vspace{1.5mm}
        17&  $1.04$& $1.20$& $1.44$& $1.72$& $2.07$& $2.50$& $3.01$& $3.50$& $3.92$& $4.40$& $4.57$& $4.75$& $4.93$& $5.13$& $5.33$& $5.54$& $5.76$& $5.99$ \\ 
\vspace{1.5mm}
        18&  $1.04$& $1.20$& $1.44$& $1.74$& $2.09$& $2.53$& $3.06$& $3.56$& $4.00$& $4.49$& $4.67$& $4.85$& $5.04$& $5.24$& $5.45$& $5.67$& $5.90$& $6.13$ \\ 
\vspace{1.5mm}
        19&  $1.04$& $1.20$& $1.45$& $1.75$& $2.11$& $2.56$& $3.10$& $3.62$& $4.07$& $4.58$& $4.76$& $4.95$& $5.15$& $5.35$& $5.57$& $5.79$& $6.03$& $6.27$ \\ 
\vspace{1.5mm}
        20&  $1.04$& $1.21$& $1.46$& $1.76$& $2.13$& $2.59$& $3.14$& $3.67$& $4.13$& $4.66$& $4.85$& $5.04$& $5.24$& $5.46$& $5.68$& $5.91$& $6.16$& $6.41$ \\
\vspace{3mm}
\enddata
\tablecomments{}
\label{Tab:StatVR}
\end{deluxetable*}
%

\section{Estimate of the areal density in a non uniform beam}
\label{Ap:Rate}

Typically, the sensitivity of a radio telescope is not uniform
across its field of view, and depends on the radial angular distance
from the beam center.
In order to convert areal density, $\kappa_{0}$, of sources brighter
than flux density $f_{0}$, to the expected number
of detectable events by a radio telescope
we need to take into account the beam pattern and
the source number count function.

We parametrize
the cumulative density of events as a function of flux
as a power law
\begin{equation}
\kappa(>f) \equiv \int_{f}^{\infty}{\kappa(f)df} = \kappa_{0} (f/f_{0})^{-\alpha},
\label{Eq:logNlogS}
\end{equation}
where 
$\kappa(f)df$ is the number of sources per flux density interval,
and $\alpha$ is the power-law index of the source
number count function.
For a uniform density population with arbitrary luminosity function
in an Euclidean universe $\alpha=3/2$.

The number of sources that can be detected in a single beam in a single epoch
up to an angular distance $r_{{\rm max}}$ from the beam center is
\begin{eqnarray}
N_{{\rm b}} & = & \int_{0}^{r_{{\rm max}}}{2\pi r dr}
              \int_{f_{{\rm min}}(r)}^{\infty}{\kappa(f)df} \cr
            & = & \int_{0}^{r_{{\rm max}}}{2\pi r \kappa_{0} [f_{{\rm min}}(r)/f_{0}]^{-\alpha}dr},
\label{Eq:Nb}
\end{eqnarray}
where $r$ is the distance from the beam center,
$f$ is the flux density,
and $f_{{\rm min}}(r)$ is the detection threshold as a function
of angular distance $r$.
For convenient we will assume that the beam pattern is Gaussian so that
\begin{equation}
f_{{\rm min}}(r) = f_{{\rm min}, 0} e^{+r^{2} \ln{2}/(r_{{\rm HP}}^{2})},
\label{Eq:Gaussian}
\end{equation}
where $r_{{\rm HP}}$ is the half width at half
power\footnote{Related to $\sigma$ of the Gaussian by $r_{{\rm HP}}=\sigma\sqrt{2\ln{2}}$},
and $f_{{\rm min}, 0}$ is the detection limit at the beam center (i.e., $r=0$).

In the case of $\alpha=3/2$ and a Gaussian beam pattern,
the integral in Equation~\ref{Eq:Nb} has an analytic solution
\begin{equation}
N_{{\rm b}}(\alpha=3/2) = -\frac{2\pi \kappa_{0} r_{{\rm HP}}^{2}}{3\ln{2}}
               \Big[ \frac{f_{{\rm min}, 0}}{f_{0}} e^{+r^{2} \ln{2}/r_{{\rm HP}}^{2}} \Big]^{-3/2} \Big|_{0}^{r_{{\rm max}}}.
\label{Eq:NbUni}
\end{equation}
For other values of $\alpha$ this integral can be evaluated numerically.
Re-arranging Equation~\ref{Eq:NbUni},
the surface density, assuming $\alpha=3/2$, is given by
\begin{equation}
\kappa_{0} = \frac{3N_{b}\ln{2}}{2\pi r_{HP}^{2}}
             \Big( \frac{f_{min,0}}{f_{0}}\Big)^{3/2} 
             \Big(1 - e^{-3 r_{max}^{2} \ln{2}/(2 r_{{\rm HP}}^{2})}\Big)^{-1}.
\label{Eq:kappa}
\end{equation}
Finally, the ratio between $\pi r_{{\rm max}}^{2}$ and $N_{{\rm}}(\alpha=3/2; r_{{\rm max}})$
gives the correction factor of 1.61 we used in \S\ref{Prev}.

\end{document}